\documentclass[lettersize,journal]{IEEEtran}
\usepackage{amsmath,amsfonts}
\usepackage{algorithmic}
\usepackage{algorithm}
\usepackage{array}
\usepackage{textcomp}
\usepackage{stfloats}
\usepackage{url}
\usepackage{verbatim}
\usepackage{graphicx}
\usepackage{cite}
\usepackage{hyperref}           
\usepackage{braket}             
\usepackage{amsthm}             
\usepackage{subcaption}

\usepackage{nicematrix} 
\usepackage{tikz}
\usepackage{xcolor}
\usepackage{tikz-network}
\usepackage{adjustbox}
\usepackage{comment}
\usepackage{booktabs}
\usepackage{tikz}
\usetikzlibrary{positioning, shapes.multipart, arrows.meta, calc}
\usetikzlibrary{quantikz2}
\usepackage[most]{tcolorbox}

\usepackage[mathscr]{euscript} 

\usetikzlibrary{fit,arrows,matrix}
\newcommand\addvmargin[1]{
  \node[fit=(current bounding box),inner ysep=#1,inner xsep=0]{};
}
\definecolor{myBlue}{RGB}{0,51,101}
\definecolor{myNeighborSet}{HTML}{005B5F}
\definecolor{myTrackingSet}{HTML}{735D9F}
\definecolor{myPurple}{RGB}{158,57,106}
\definecolor{darkblue}{HTML}{1f4e79}
\newcounter{boxcounter}

\newcommand{\eqdef}{\stackrel{\triangle}{=}}

\tcbuselibrary{skins}

\newcommand{\stepboxcounter}[1]{%
  \refstepcounter{boxcounter}%
  \label{#1}%
}

\newtheorem{lem}{Lemma}

\newtheorem{proper}{Property}

\theoremstyle{remark}

\theoremstyle{definition}
\newtheorem*{contrib}{Contributions}
\newtheorem{defin}{Definition}
\newtheorem{design}{Design Principle}
\newtheorem{subdesign}{Design Principle}

\newtheorem{remark}{Remark}

\begin{document}

\title{Quantum Internet Architecture: unlocking Quantum-Native Routing via Quantum Addressing}
\author{Marcello Caleffi \textit{and} Angela Sara Cacciapuoti \\
    \textsc{Invited Paper}
    \thanks{M. Caleffi and A.S. Cacciapuoti are with the \href{www.quantuminternet.it}{www.QuantumInternet.it} research group, University of Naples Federico II, Naples, 80125 Italy. E-mail: \href{mailto:angelasara.cacciapuoti@unina.it}{angelasara.cacciapuoti@unina.it},  \href{mailto:marcello.caleffi@unina.it}{marcello.caleffi@unina.it}.}
    \thanks{This work has been funded by the European Union under the ERC grant QNattyNet, n.101169850. Views and opinions expressed are however those of the author(s) only and do not necessarily reflect those of the European Union or the European Research Council. Neither the European Union nor the granting authority can be held responsible for them.}
}

\maketitle

\begin{abstract}
The key objective of the Quantum Internet is the distribution and manipulation of entanglement to enable unprecedented applications. This requires a radical departure from classical Internet design principles, such as the end-to-end argument, due to the inherently stateful and non-local nature of entanglement, which demands coordinated in-network operations and persistent state awareness. To this end, we propose a novel hierarchical Quantum Internet architecture centered on the concept of \textit{Entanglement-Defined Controller} (EDC). This architectural design constitutes the foundational layer, by enabling a clear separation between control and data planes. While necessary, this separation is insufficient to manage entanglement resources, requiring a \textit{quantum-native control plane}. Consequently, we propose a \textit{quantum addressing scheme} that embeds quantumness directly into node identifiers, allowing the network to natively track and manipulate entanglement as a dynamic resource. Built upon these two interdependent pillars -- EDC-based architecture and quantum addressing -- we design a \textit{quantum-native routing protocol} that achieves scalability through compact routing tables, by efficiently operating over entanglement-defined topologies. Finally, we design a \textit{quantum address splitting} functionality based on Schr\"odinger's oracles that generalizes classical match-and-forward logic to the quantum domain. Collectively, these contributions demonstrate, for the first time, the fundamental advantages of quantum-by-design network control for enabling scalable quantum networking.
\end{abstract}


\section{Introduction}
\label{sec:1}
The Quantum Internet \cite{CacCalTaf-20, CacCalVan-20,rfc9340,Kim-08,VanMet-14,DurLamHeu-17} promises unprecedented capabilities, including unconditionally secure communication, distributed quantum computing, and enhanced sensing \cite{rfc9583}. At the core of these capabilities lies quantum entanglement, the fundamental communication resource of the Quantum Internet \cite{rfc9340}, which demands a radical architectural departure from the classical Internet design principles \cite{CacIllCal-23,IllCalMan-22}. 

Specifically, the end-to-end principle \cite{rfc1958} breaks down in the Quantum Internet, since the network paradigm shifts from transmitting information to distributing and managing entangled states \cite{CacCalTaf-20,IllCalMan-22}. Unlike classical bits, entanglement generation, distribution, storage and exploitation inherently require in-network operations and the maintenance of related information inside the network. To elaborate more, classical bits are inherently stateless: intermediate network nodes can process and forward them, without the need to retain any additional information or detail about the state of the end-to-end communication. In contrast, the temporal constraints imposed by decoherence, along with the inherent complex mechanisms underlying the generation and maintenance of quantum entanglement, necessitate that the network nodes retain \textit{state information}. As a pivotal example, nodes must be aware of the residual coherence time of the stored entangled qubits (e-bits), for properly operating on them. Hence, e-bits are
fundamentally \textit{stateful} \cite{IllCalMan-22}, directly contradicting the classical end-to-end argument proposed by Saltzer in \cite{SalReeCla-84}, outlining the classical stateless design.

Furthermore, the \textit{non-local nature} of quantum entanglement requires additional state information for its effective exploitation, beyond simply identifying the nodes that first shared it. Indeed, in EPR-based networks, entanglement can be swapped, by changing at run-time the identities of the entangled nodes. In more complex scenarios where multipartite entanglement is exploited, the above dynamism is further enriched. In fact, multipartite entanglement can be manipulated and reconfigured across subsets of network nodes, namely, across entire sub-networks. Accordingly, since entanglement is not information per-se but rather a communication resource \cite{rfc9340}, its value and utility extend well beyond the original source-destination pair or the original initiating sub-network. If left uncoordinated, these non-local effects can trigger the so-called amplification principle \cite{rfc3439}, where uncontrolled entanglement resources lead to routing ambiguities, resource inefficiencies, and ultimately network instability, thereby undermining scalability. Therefore, effective tracking and management of entanglement resources are essential for scalable quantum network architectures.

\begin{figure*}[!t]
    \centering
    \includegraphics[width=7in]{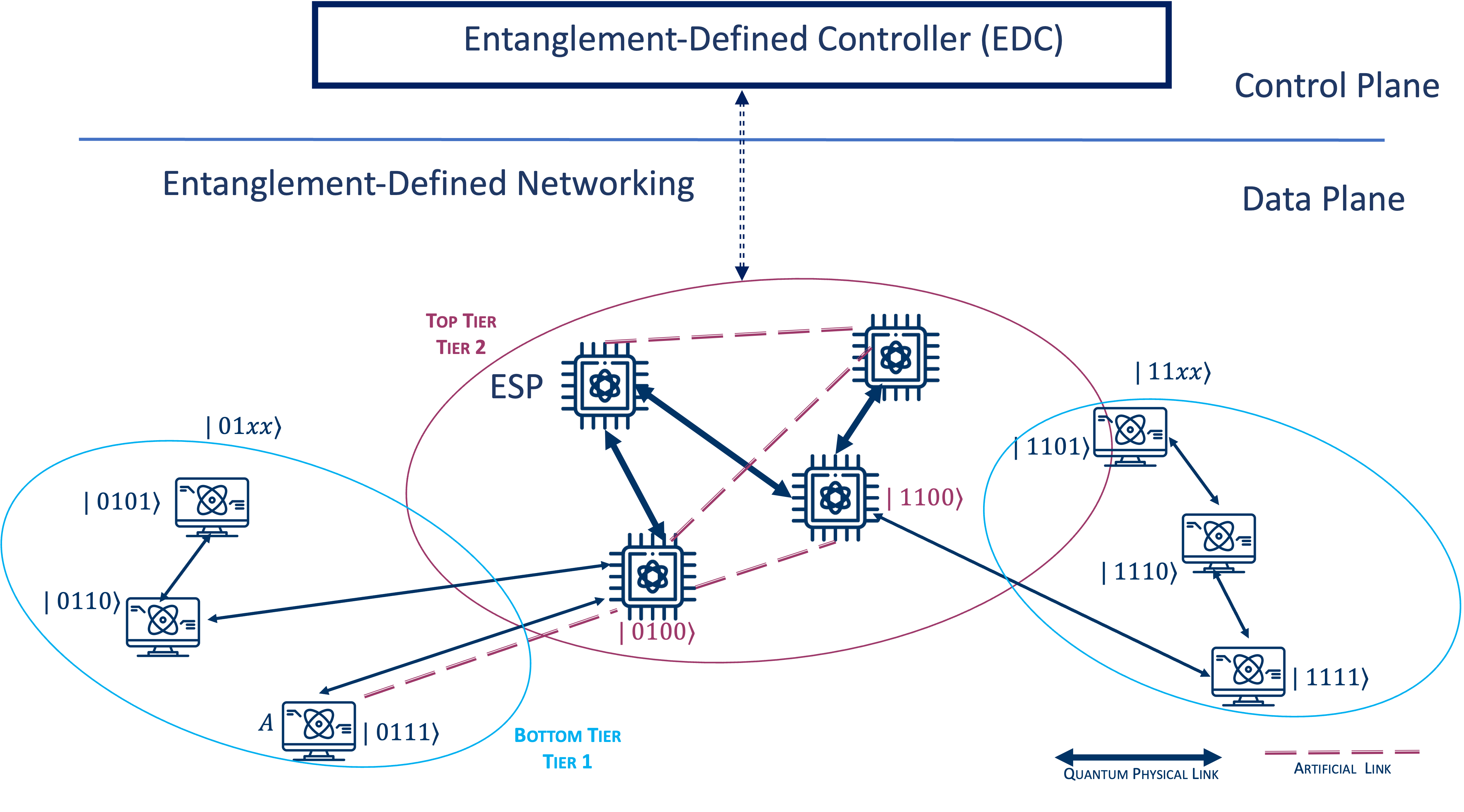}
    \caption{Entanglement-Defined Network Architecture showing: (1) ESPs forming a virtual mesh via proactive entanglement sharing (dashed lines), (2) EDC responsible of the control plane functionalities, and (3) end-user quantum nodes connected to the serving ESP.}
    \label{fig:01}
    \hrulefill
\end{figure*}

Building on the above considerations, to efficiently manage the in-network entanglement operations while at the same time preserving the \textit{simplicity principle} that shaped classical Internet design, we propose a novel architectural framework, which integrates \textit{quantum-native} functionalities at its core. 
Specifically, by extending the architectural vision pioneered in \cite{CacIllCal-23}, this paper formalizes a two-tier Quantum Internet architecture as depicted in Fig.~\ref{fig:01}, which centralizes control of entanglement operations through an \textit{Entanglement-Defined Controller (EDC)}. Analogous to the SDN controller in classical networks, the EDC coordinates in-network operations and supports scalable entanglement management, as further detailed in Sec.~\ref{sec:2}.   

The aforementioned EDC-based architecture is at the core of our proposal, since it enables a clear separation between control and data planes -- a necessary condition for scalability. However, this architectural separation alone is not sufficient to
manage the coordinated in-network operations and persistent state awareness, required by entanglement. Crucially, scalability also depends on another dimension: how the \textit{control plane} itself is designed \cite{rfc9340}. Specifically, the control plane must embrace quantum principles and phenomena to manage entanglement dynamics effectively. The rationale behind this claim stems again from the non-locality of quantum entanglement: entanglement proximity cannot be confined to physical locality or restricted to fixed neighborhoods. This has a profound implication: \textit{entanglement redefines the very same concept of topological neighborhood} \cite{CacIllCal-23,CheIllCac-24,MazCalCac-25}. 

As a result, a conventional control plane, built upon classical assumptions of locality and IP-like addressing, cannot timely track, respond to, or propagate entanglement state changes across the network. Without a fundamental rethinking of network addressing and control mechanisms, the control plane itself becomes inevitably the bottleneck to scalability. This issue is not entirely unfamiliar: even in classical networks where entanglement is absent, it has been shown that the number of control messages per topology change, namely, the updating communication overhead, cannot scale better than linearly on Internet-like topologies \cite{KriClaFal-07}. In the quantum setting, the challenge is even more severe due to the intrinsic statefulness and fragility of entanglement.

Stemming from the above, we complement the proposed architecture with a novel addressing scheme that embeds quantum principles and phenomena directly into the node identifiers, thereby elevating \textit{the control plane to a quantum-native level}. Indeed, it is the proposed addressing scheme that makes possible to operate over entanglement-defined topologies with compact, dynamic, and scalable routing tables. More in detail, we design a \textit{quantum addressing scheme} that leverages quantum superposition. This enables a paradigmatic departure from the underlying assumption beyond classical IP addressing: \textit{a sequence of (32/128) bits encoding a single network address}, reflecting the node location within the physical network topology. In fact, a sequence of (say $n$) qubits can encode a single node identity, i.e., a single quantum network address, but it can also encode a superposition of node identities, i.e., a superposition of multiple quantum states with each state denoting a network address. Thus, a sequence of $n$ qubits can represent a set of quantum nodes. Crucially and completely differently from classical IP subnet addresses, a quantum network address can represent a set of quantum nodes regardless of their physical position within the quantum network. 

To concretely demonstrate that the quantum-native functioning unlocked by the foundational layer -- composed of the two interconnected pillars, the EDC-based architecture and quantum addressing -- enables scalability, we design two different \textit{quantum-native routing protocols}. These protocols exploit superposed quantum addresses to achieve compactness, namely sublinear routing table sizes. A detailed analysis is provided in Sec.~\ref{sec:3}, where we formally demonstrate the capability of the design protocols to operate efficiently over entanglement-defined topologies. 

In this light, it becomes crucial to allow a node to extract and act on individual node identities within a superposed address. Consequently, we design a \textit{quantum addressing splitting functionality} in Sec.~\ref{sec:4}, which generalizes the classical match-and-forward paradigm of classical networks to the quantum domain. The proposed quantum addressing splitting has been conceived by modifying Grover’s quantum search algorithm in order to tailor quantum-superposed data structures. In our proposal, the oracle is coherently controlled by quantum sub-network addresses, so that the presence or the absence of phase inversion becomes a controllable quantum degree of freedom.\\

\begin{figure*}[!t]
    \centering
    \includegraphics[width=5.5in]{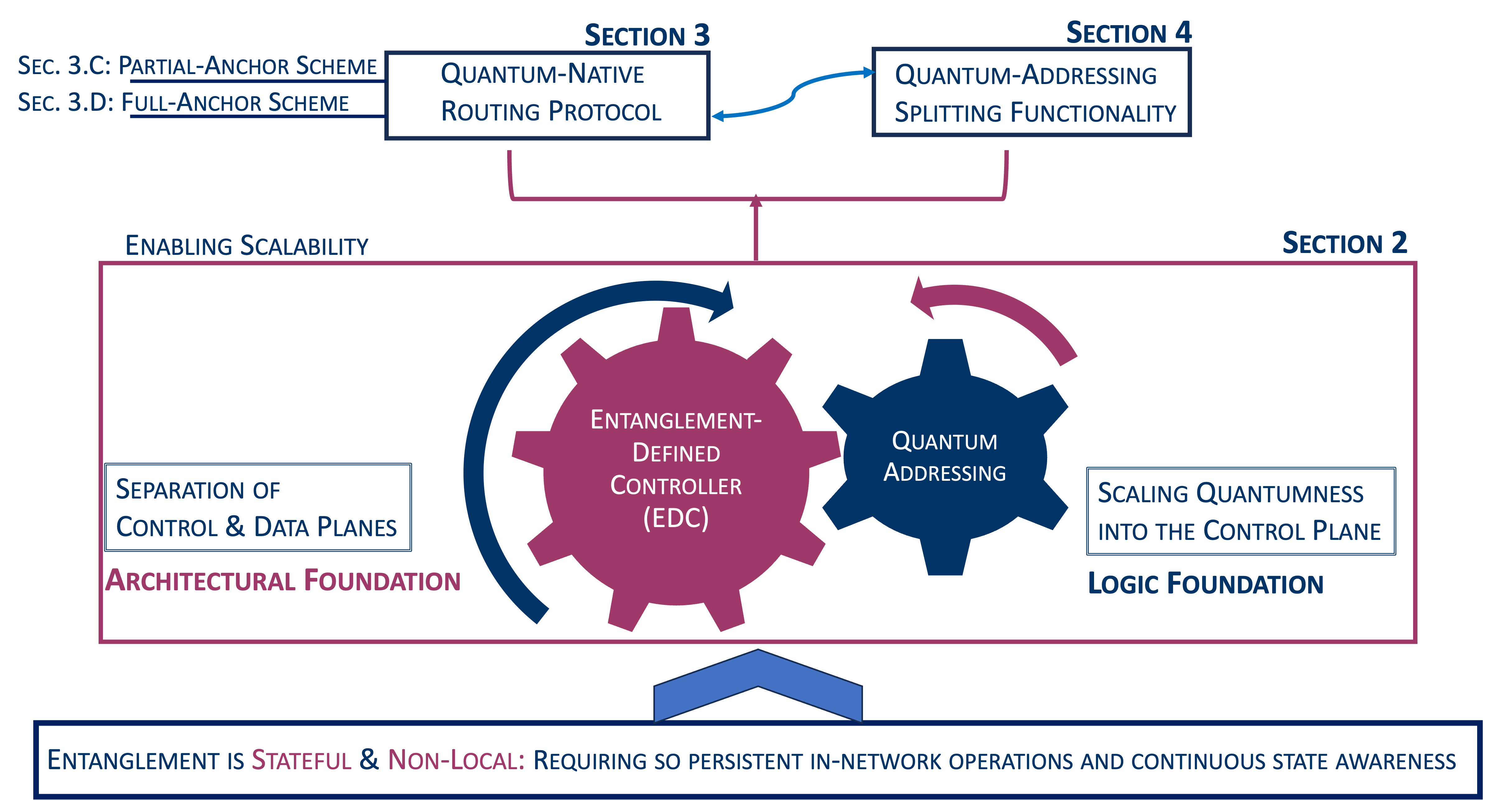}
    \caption{Overview of the contributions and their dependencies. The EDC-based architecture and quantum addressing form the two interdependent pillars of the foundational layer of our proposal, enabling scalable quantum-native control. Building on this core, the routing protocol and the addressing splitting functionality demonstrate concrete scalability gains.}
    \label{fig:01_bis}
    \hrulefill
\end{figure*}

\begin{contrib}
    To sum up and as represented in Fig.~\ref{fig:01_bis}, the contributions of this paper are fourfold:
    \begin{itemize}
        \item the design of an architectural framework to manage in-network entanglement operations efficiently, centered around the Entanglement-Defined Controller (EDC);
        \item the design of a \textit{quantum addressing scheme}, which encodes quantum properties and behaviors directly into node identifiers, by scaling the control plane to a quantum-native level;
        \item the design of \textit{quantum-native} routing protocols that fully exploit the quantum nature of the addressing scheme to achieve scalable and compact routing tables;
        \item the proposal of a \textit{quantum addressing splitting functionality} that extends classical forwarding operations to quantum-superposed identifiers,  through the design of a Schr\"odinger's oracle for Grover's search algorithm, coherently controlled by quantum addresses.
    \end{itemize}
\end{contrib}

To the best of our knowledge, this is the first work demonstrating with concrete architectural and protocol design the advantages of the quantum-native functioning of the network, that was previously only anticipated at a conceptual level in \cite{CacIllCal-23}.

The remaining part of the paper is organized as follows. In Sec.~\ref{sec:2}, we detail the two-tier architecture, by clarifying and substantiating some key deviations from the classical Internet design principles. In Sec.~\ref{sec:3}, we design the quantum-native routing protocols, built upon the quantum addressing scheme. In Sec.~\ref{sec:4}, we propose the quantum addressing splitting functionality. Finally, in Sec.~\ref{sec:5} we discuss some key aspects of the proposal along with future directions. 

\section{Quantum Internet Architecture}
\label{sec:2}

In this section, we propose some architectural principles for the Quantum Internet design, which should not be interpreted as dogmas but rather as pragmatic guidelines and criteria for harvesting the unique properties of quantum entanglement. Our design perspective, while departing from the classical Internet, aligns with a fundamental insight from its evolution, articulated in RFC 1958 \cite{rfc1958}:
\begin{quote}
    ``\textit{the principle of constant change is perhaps the only principle of the Internet that should survive indefinitely}''. 
\end{quote}
We argue this is one of the most valuable lessons from the classical Internet evolution.

\subsection{Internet Architecture in a nutshell}
\label{sec:2.1}
By oversimplifying, in the classical Internet, Internet Service Providers (ISPs) form a multi-tiered hierarchy where lower-tier ISPs connect to higher-tier ones, creating a mesh of packet-switched networks, that provide end-to-end connectivity to end users. This hierarchy design has been the result of a long and largely unforeseen evolution, where a complex interplay of technical and economic factors played a crucial role. In this context, the \textit{end-to-end principle} \cite{rfc1958} emerged as a key architectural tenet. Accordingly, end-to-end protocol design should not rely on state maintenance inside the network, i.e., on information about the state of the end-to-end communication. Rather, such a state should be maintained only at the end points, so that the state can be destroyed only if the end point itself breaks \cite{rfc3439}. The overall effect\footnote{
It is worthwhile to observe that this observation is not intended to imply a strict causal relationships between design principles and historical evolution. The architectural foundations of the Internet were profoundly shaped by a confluence of factors, including specific technical choices and priorities.  
For instance, ``survivability'' is identified in \cite{Cla-88} as a core design objective and the essential rationale behind the end-to-end principle (the “fate-sharing’’ concept in \cite{Cla-88}). At the same time, as pointed out to us by John Day, the origin of the end-to-end principle has to be traced to CYCLADES, the French project led by Louis Pouzin, which also introduced the datagram. CYCLADES explicitly recognized that the network only had to provide ``best-effort'' delivery, since hosts would never assume it was reliable and would instead ensure end-to-end reliability themselves. This insight implied that the network could be faster, simpler, and less costly, and laid down fundamental principles that were later incorporated into the Internet architecture \cite{Day-08}.} 
of the \textit{end-to-end principle}, coupled with the \textit{simplicity principle} \cite{rfc1958}, resulted in the classical Internet architecture, where intelligence is localized at the network edges rather than hidden inside the core network. In other words, Internet has smart edges where applications and operating systems reside and provide complex communication functionalities, and a simple core, consisting of stateless packet-forwarding engines and a control plane offering mainly best-effort datagram delivery \cite{rfc3439}.

\subsection{The need for Entanglement-Packet Switching}
\label{sec:2.2}

The aforementioned design principles -- centered around stateless routing, best-effort delivery, and end-to-end principle -- are fundamentally inadequate for the Quantum Internet, which requires a radical departure from classical networking paradigms. This shift fundamentally transforms the core network operations, as detailed in the following.

First, a preliminary consideration must be introduced: informational qubits cannot be simply adapted to packet-switching paradigm, due to the fundamental constraints imposed by the no-cloning theorem and the quantum measurement postulate. In fact, unknown qubits cannot be perfectly copied or amplified, and any measurement irreversibly alters their state. This prohibits the adoption of conventional store-and-forward packet-switching paradigm, and it makes traditional best-effort delivery, where packets may be lost and retransmitted, inherently incompatible with quantum information transfer. 

Entanglement distribution circumvents these limitations, by establishing quantum correlations as the fundamental network resource, allowing us to move beyond the direct transmission of informational qubits \cite{CacCalVan-20,CacCalTaf-20}. Indeed, since entanglement is a communication resource rather than information itself, it is not constrained by the no-cloning theorem \cite{IllCalMan-22}. And, through the quantum teleportation protocol \cite{CacCalVan-20}, pre-shared entanglement enables reliable quantum information transfer, without the need of physically transmitting an information carrier.

From the above, it follows that the Quantum Internet must embrace a fundamentally new paradigm:
\begin{quote}
    \textit{ebits serve as the basic network ``packets'', carrying quantum correlations across network nodes. We term this paradigm as entanglement-packet switching.}
\end{quote}

This transition is not merely an optimization or a modification of the current packet-switching paradigm, but a fundamental departure, imposed by the unique constraints of quantum mechanics:
\begin{quote}
\textit{While the goal of the classical packet-switching paradigm is to determine the best next-hops toward a set of nodes (routing) and to forward packets through these next-hops from source to destination, entanglement-packet switching aims to distribute and manipulate entanglement among quantum nodes, ultimately entangling the source and destination regardless of their physical location.}
\end{quote}
Accordingly, this paradigm shift enables the following two key features.
\begin{itemize}
    \item{\textit{Decoupled design:}} entanglement-packet switching fully \textit{decouples} the quantum information transfer from the physical transmission of information carriers, overcoming the limitations imposed by the no-cloning theorem and the quantum measurement postulate on the direct informational-qubit transmission.
    \item{\textit{Scalability and compatibility:}} entanglement-packet switching \textit{retains} the flexibility and inherent scalability of packet-switching architectures, while ensuring backward compatibility with the classical Internet \cite{DiAQiMil22,YooSinKum24,VisHolDia24}. Hence, the approach is at the same time forward-looking and backward-compatible.
\end{itemize}

Given this fundamental transition towards entanglement-packet switching, we can now better digest the rationale for the incompatibility between the end-to-end principle (as said, substratum of the classical Internet design) and the Quantum Internet, highlighted in Sec.~\ref{sec:1}. This breakdown stems directly from the unique properties of quantum entanglement. 

Indeed, as described in Sec.~\ref{sec:1}, the complex, challenging stateful nature of entanglement along with its non-local effects call for \textit{in-network operations and persistent state awareness} across all phases of the entanglement life-cycle -- from generation through distribution to storage and final utilization -- by directly contradicting the classical end-to-end argument. 

\begin{table}[!t]
    \centering
    \renewcommand{\arraystretch}{1.4}
    \begin{tabular}{p{0.45\columnwidth}|p{0.45\columnwidth}}
    \hline
        \hline
        \textbf{Classical Internet} & \textbf{Quantum Internet} \\
        \hline
        complexity located at the network edges & complexity concentrated in the core network \\
        \hline
        stateless core network & stateful core network \\
        \hline
        end-to-end protocol design & network-mediated protocol design \\
        \hline
        \hline
    \end{tabular}
    \caption{Classical vs Quantum Internet architecture: concise comparative overview.}
    \label{tab:01}
    \hrulefill
\end{table}

Accordingly, the Quantum Internet demands a complete inversion of the classical design philosophy, as summarized in Table~\ref{tab:01}.

\begin{remark}
    Indeed, when the above considerations are combined with the sophisticated and resource-intensive setups required by the state-of-the-art hardware, it becomes both practical and efficient to advocate for concentrating the complexity inside the core network, while leaving the edges of the network simpler. 
\end{remark}

\subsection{The Two-Tier Hierarchy}
\label{sec:2.3}

Building on the above considerations, we propose a two-tier Quantum Internet architecture, by distinguishing between \textit{entanglement service providers (ESPs)} and \textit{quantum edge nodes}, as represented in Fig.~\ref{fig:01}.
\begin{itemize}
    \item{\textit{Top Tier}:} ESPs act as the highest tier,  referred to also as tier-2, analogous to ISPs in the classical Internet. They form the ``entangled-backbone''\footnote{It is worth emphasizing that, in the classical Internet, the terms ``core network'' and ``backbone network'' are often used interchangeably, despite a subtle, yet important, distinction between them. In this paper, we adopt the same interchangeable usage in the context of the Quantum Internet, where these terms are borrowed primarily by analogy. Indeed, given that the Quantum Internet remains in its infancy and early conceptual stages, a formal distinction between the two has yet to emerge or become necessary.}, by providing end-to-end entanglement connectivity to the lowest tier, via proactively maintenance of entangled resources. The EPSs are interconnected via long-distance quantum links, such as optical fibers and they are equipped with the sophisticated and resource-intensive infrastructure required for entanglement generation and distribution. 
    \item{\textit{Bottom Tier}:} edge quantum nodes, which include quantum processors, sensors, cryptographic devices, acting as the edge tier (referred to also as tier-1) consuming entanglement resources to fulfill the quantum applications needs. They are mainly connected to the nearest ESP via short-range quantum links.
\end{itemize}

\begin{remark}
    The two-level hierarchy introduced so far is not intended to represent a definitive architecture for the Quantum Internet, which is still in its early stage conceptualization\footnote{It is likely that the Quantum Internet will evolve into a multi-level hierarchy of increasing complexity, as happened to the classical Internet.}. Rather, it serves as a reference model for capturing the distinguishing features of quantum entanglement in contrast to classical information.
\end{remark}

In the remaining part of this manuscript we focus on the design of the \textit{top tier}, given its pivotal role in enabling large-scale deployment of the Quantum Internet. 

To this aim, we must first recall a key point from a networking perspective. Specifically, it has been extensively shown in literature that entanglement, once shared, enables a new and richer form of connectivity, with no-counterpart in classical networks. The activated entanglement-proximity gives rise to an overlay topology, often referred to as \textit{artificial topology}, established upon the physical topology \cite{IllCalMan-22,MazCalCac-25,CheIllCac-24,RamDur-20}. And differently from the physical topology, the entanglement-enabled overlay is inherently dynamic\footnote{We refer the reader to \cite{CacIllCal-23,IllCalMan-22} for an in-depth discussion about the artificial topology dynamics, which differ profoundly from the ones induced by node mobility in classical networks.}, 
since entanglement can be locally manipulated while still producing global coverage effects, due to its inherently non-local nature. Thus, the artificial topology can be reconfigured on-the-fly via local quantum operations (with entanglement swapping the archetypal example), by enabling global connectivity changes without modifying physical links. This adaptability is essential to accommodate evolving end-to-end entanglement requests, reduce latency -- key in scenarios very sensitive to decoherence -- and mitigate the limitations imposed by the physical topology \cite{MazCalCac-25}. 

In this perspective, we envision the ESPs to proactively establish and maintain entanglement resources among each others, by exploiting the underlying physical network topology. The activated artificial topology serves as the functional equivalent of the physical mesh network interconnecting ISPs. Crucially, this topology is continuously refreshed and reconfigured to reflect the current status and availability of entangled resources, thereby enhancing the overall reliability, adaptability, and robustness of the network.  Moreover, the proactive nature of the strategy allows to face with the non-persistence of entangled resources, since the Quantum Internet must operate under the constraint of entanglement depletion, upon use. The overall result is a dynamic artificial mesh, where links are continually created and consumed, in response to quantum application demands.

\subsection{Entanglement-Defined Controller}
\label{sec:2.4}

To efficiently manage in-network operations and state maintenance while addressing scalability, an Entanglement-Defined Controller (EDC) orchestrates the entanglement resources among the ESPs, by mirroring the role of a Software-Defined Networking (SDN) controller in classical architectures \cite{KreRamVer-14}. The EDC oversees three main crucial functions: 
\begin{itemize}
    \item \textit{Reconfiguration:} the dynamic management and reconfiguration of entangled resources among ESPs;
    \item \textit{Monitoring:} the monitoring of the fidelity and availability of entanglement resources across ESPs;
    \item \textit{Policy enforcement:} the enforcement of global policies for routing, resource allocation, and entanglement loss recovery.
\end{itemize}

By centralizing the control logic within the EDC, we lay the foundation for a scalable, adaptable, and programmable Quantum Internet architecture, necessary to manage quantum entanglement as a dynamic, non-persistent network resource. Indeed, the EDC\footnote{Please refer to Sec.~\ref{sec:5}, in which we provide further details on the EDC, by including how the proposed architecture can support multiple, potentially federated EDCs rather than relying on a single controller.} enables coordinated management of entangled resources across the network, thereby supporting the inherent network-mediated nature of quantum communication protocols. Moreover, the EDC enables a dynamic response to evolving application-demands and to the intrinsic volatility of entangled resources. 
It is important to highlight that, for maintaining the overlay topology, we do not force any EDC to have a persistent global network knowledge. In fact, in Sec.~\ref{sec:4}, we design a quantum-native routing protocol that operates effectively under limited or local network visibility.

\begin{table*}[t]
    \centering
    \renewcommand{\arraystretch}{1.4}
    \begin{tabular}{p{0.14\linewidth}|p{0.37\linewidth}|p{0.37\linewidth}}
        \hline
        \hline
        \textbf{network feature} & \textbf{classical Internet} & \textbf{Quantum Internet} \\
        \hline
        \textbf{resource persistence} & communication links are permanent, i.e., their dynamics largely exceed the forwarding dynamics & entanglement is ephemeral and depleted upon use \\
        \hline
        \textbf{control plane} & populates routing tables with best hops toward destinations & encodes routing information via quantum superposition exploiting the quantum addressing scheme and orchestrates entanglement resources so that they are efficiently distributed \\ & &
         \\
        \hline
        \textbf{data plane} & packet forwarding & quantum operations and entanglement-packet forwarding \\
        \hline
        \hline
    \end{tabular}
    \caption{Control and data plane: classical Internet vs the proposed Quantum Internet architecture.}
    \label{tab:02}
    \hrulefill
\end{table*}

Stemming from the above, the proposed EDC-based architecture establishes a clear separation between
\textit{control plane} and \textit{data plane} -- a necessary condition for scalability. In this architectural separation, the EDC is responsible for control-plane functionalities, while the ESPs handle data plane functionalities, as summarized in Table~\ref{tab:02}.

As for the control plane, it is important to note that the quantum routing literature is already very extensive, with significant progress made in the field. Representative works include \cite{Cal-17,LeNgu22,CheXueLi-24,AbaCubMai-25}. However, a common trait of these approaches is their reliance on classical communications for the control-plane functionalities. This aspect is thoroughly analyzed and highlighted in the most recent and comprehensive survey by Abane \textit{et al.}~\cite{AbaCubMai-25}. Consequently, network control in the state-of-the-art solutions remains completely rooted in the classical infrastructure, which, as discussed in Sec.~\ref{sec:1}, poses fundamental scalability limitations. Conversely, our work departs from this paradigm by advocating for \textit{quantum-native control functionalities}, thereby unlocking scalability at the architectural level.

More in detail, we design a \textit{quantum control plane}, where routing information is encoded into quantum states, by leveraging the quantum addressing scheme designed in Sec.~\ref{sec:2.6}. These ``quantum-encoded routes'' are then stored within quantum routing tables and manipulated through quantum operations for the forwarding. This approach allows scalable and efficient path selection, without requiring persistent global network knowledge, as shown in Sec.~\ref{sec:3}.

In this way, \textit{we scale the network to a quantum-native functioning}. 

As for the quantum data plane, it generalizes the classical notion of forwarding to the quantum domain, as highlighted in \cite{AbaCubMai-25}. However, to account for quantum-native functioning, key extensions beyond the discussion in \cite{AbaCubMai-25} must be introduced.

Specifically, in classical networks, the forwarding logic follows a match-and-forward paradigm, where the destination address is extracted from the packet header and matched against the routing table. In our architecture, this paradigm is extended along three dimensions: 
\begin{itemize}
    \item \textit{Entanglement-packet forwarding via quantum addressing:} ESP identifiers are encoded into quantum states via quantum addresses, as described in the next subsection. Thus, \textit{forwarding decisions} require the capability to operate directly on quantum identifiers. This is enabled by a quantum header in the packet, as illustrated in Fig.~\ref{fig:02}, which carries the quantum equivalent of source-destination addresses. This approach generalizes ``next-hop'' selection to the quantum realm, by applying appropriate quantum operations to quantum addresses in routing tables, thereby eliminating classical header parsing. This mechanism is detailed in Secs.~\ref{sec:3} and \ref{sec:4}.
    \item{\textit{End-to-end entanglement establishment:}} The data plane actively supports the establishment of end-to-end (E2E) entanglement across ESPs for the bottom-tier demands. This involves generating elementary hop-by-hop entangled links, which may not yet exist at the time of the request, and performing quantum operations, such as entanglement swapping and purification. Consequently, quantum packets, which carry ebits, are propagated through sequences of quantum operations, rather than merely through transmission over physical links.
    \item{\textit{Artificial topology maintenance:}} The maintenance of the artificial topology among the ESPs requires the data plane to continuously regenerate virtual links depleted during forwarding. This is achieved through elementary entanglement-link generation, quantum operations and refreshing of entangled resources.
\end{itemize}

This separation between data and control plane ensures scalability while facing with the ephemeral nature of quantum entanglement. For instance, a bottom-tier request is fulfilled, by manipulating and hence by reconfiguring the overlay among ESPs (e.g., stitching virtual links via swaps), while the control plane concurrently instructs the data plane to repair the overlay, by redistributing elementary entanglement. This maintains the topological robustness. Thus, while the data plane consumes pre-established entanglement for applications (e.g., teleportation), the control plane continuously refills depleted resources.

\subsection{Quantum Addressing Scheme}
\label{sec:2.5}
As aforementioned, we scale the network to a quantum-native functioning, by designing control plane functionalities via quantum state manipulations. To this aim, we embrace quantumness within the node addresses \cite{CacIllCal-23}. It is worthwhile to note that the quantum addressing is not a substitute of the classical network addressing. 
Indeed, each network node must be equipped with two types of addresses: i) a classical address, such as an IP address, which is mandatory for the classical communications and signaling required by any quantum communication protocol; ii) and a quantum address, which facilitates efficient and scalable control functions.

This dual-addressing framework ensures backward compatibility with the classical infrastructure, while unlocking quantum advantages in control-plane scalability and entanglement orchestration, as proved in Sec.~\ref{sec:3}.

In assigning the quantum addresses, we hybridize classical hierarchical principles with quantum features. Specifically, each node in the network, regardless of being either an EPS or a tier-1 node, is assigned a quantum address represented by a computational basis state of an $N$-qubit system. Thus, the addresses are orthogonal quantum states, ensuring so perfect distinguishability. We account for the two-tier hierarchy introduced so far by imposing tier-1 addresses to share a common prefix with the address of the serving ESP. We envision that this addressing scheme enables efficient routing and entanglement management across the quantum network.

In Fig.~\ref{fig:01}, we provide a pictorial representation of a toy model for the above scheme, by considering a network of $n=16$ nodes with $4$ ESPs. Each ESP serves up to $3$ tier-1 nodes, and their corresponding address prefix determining the clustering are:
\begin{equation}
    \label{eq:01}
    \ket{00xx}, \ket{01xx}, \ket{10xx}, \ket{11xx}. 
\end{equation}

Formally, the quantum network is modeled as an undirected graph:
\begin{equation}
    \label{eq:02}
    G=(V,E),
\end{equation}
where:
\begin{itemize}
    \item the vertex set $V$ represents the $|V| = n$ quantum nodes, partitioned in two disjoint subsets: $V_1$, representing edge quantum (tier-1) nodes with $|V_1|=n_{u}$, and $V_2$, representing ESPs (tier-2 nodes), with $|V_2|=n_{e}$. Accordingly:
    \begin{equation}
        \label{eq:03}
        V = V_1 \cup V_2, \quad V_1 \cap V_2 = \emptyset,
    \end{equation}
    with $n=n_{u}+n_{e}$.
    \item the edge set $E \subseteq V \times V$ denotes the set of links interconnecting the quantum nodes. 
\end{itemize}

\begin{defin}[\textbf{Quantum Address}]
    \label{def:01}
    Let $\mathcal{B} \eqdef \left\{ \ket{x} \;\middle|\; x \in \{0,1\}^N \right\}$ denote the computational basis of the Hilbert space associated with an $N$-qubit system, where $N = \left\lceil \log_2 n \right\rceil$. \\
    Let $\mathcal{E} \eqdef \left\{ \ket{v_1}, \ldots, \ket{v_{n_e}} \right\} \subset \mathcal{B}$ be a set of $n_e$ distinct computational basis states. Let define $p = \left\lceil \log_2 n_e \right\rceil$ as the prefix length. For any $x \in \mathcal{B}$, we define $\text{prefix}_p(x)$ as the string consisting of the first $p$ bits of $x$, i.e.:
    \begin{equation}
        \label{eq:04}
        \text{prefix}_p:  \{0,1\}^N \rightarrow \{0,1\}^p.
    \end{equation}
    The elements of $\mathcal{E}$ are selected to satisfy the condition:
    \begin{equation}
        \label{eq:05}
        \text{prefix}_p(v_i) \neq \text{prefix}_p(v_j), \quad \text{for } i \neq j,
    \end{equation}
    ensuring that each element in $\mathcal{E}$ is assigned a unique prefix. Each ESP in $V_2$ is uniquely identified by a \emph{quantum address} $\ket{v_i}$, selected in $\mathcal{E}$:
    \begin{equation}
        \label{eq:06}
        \ket{v_i} \in \left\{ \ket{v_1}, \ldots, \ket{v_{n_e}} \right\}.
    \end{equation}
    In the following, we refer to an ESP as either $v_i$ or $\ket{v_i}$, depending on the context.\\
    Each edge (tier-1) quantum node in $V_1$ served by a given ESP $\ket{v_i}$ is assigned with a quantum address that shares the same $k$-prefix with $\ket{v_i}$, thereby preserving hierarchical coherence in the quantum address space. Formally, for each ESP in $V_2$ with address $\ket{v_i} \in \mathcal{E}$, we define its \textit{serving cluster} as:
    \begin{equation}
        \label{eq:z:07}
        \mathcal{C}_{\ket{v_i}} \eqdef \left\{ \ket{x} \in \mathcal{B} \;\middle|\; \text{prefix}_p(x) = \text{prefix}_p(v_i) \right\}.
    \end{equation}
    That is, each cluster contains all and only those computational basis states that share the first $k$ bits with $\ket{v_i}$.
    By construction\footnote{By construction, each cluster has the same size equal to $2^{N-p}-1$, by excluding the address of the serving ESP. To support heterogeneous cluster sizes while keeping the prefix length $p$ fixed, one would need to define a maximum cluster size $M_{max}$ and set the quantum address length as $N=p+\left\lceil \log_2 M_{\text{max}} \right\rceil$. This approach inevitably results in address spaces with unused quantum addresses in smaller clusters. However, exploring such generalizations is beyond the scope of this paper, which instead focuses on showing the powerful setup allowed by quantum addressing schemes.} the serving clusters are mutually disjoint and they constitute a partition of the computational basis set $\mathcal{B}$, i.e.:
    \begin{equation}
        \label{eq:08}
        \mathcal{B}=\displaystyle \cup_{i=1}^{n_e} \mathcal{C}_{\ket{v_i}} \quad \wedge \quad \mathcal{C}_{\ket{v_i}} \cap \mathcal{C}_{\ket{v_j}} = \emptyset, \; \forall \, i\neq j \in V_2.
    \end{equation}
\end{defin}

\subsection{Quantum Packet Structure}
\label{sec:2.6}

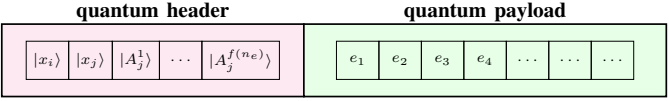
\begin{figure}[!t]
    \centering
    \begin{adjustbox}{width=\columnwidth}
    \begin{tikzpicture}
        \draw[fill=magenta!10, thick] (0,0) rectangle (5,1.2);
        \draw[fill=green!10, thick] (5,0) rectangle (11,1.2);
        \node at (2.5,1.4) {\textbf{quantum header}};
        \node at (8,1.4) {\textbf{quantum payload}};
        \node at (2.5,0.6) {
            \begin{tikzpicture}[baseline=0,MyStyle/.style={draw, minimum width=2em, minimum height=1.8em, outer sep=0pt},every node/.style={font=\scriptsize}]
                \matrix (A) [matrix of math nodes, nodes={MyStyle, anchor=center}, column sep=-\pgflinewidth]
        { \ket{x_i} & \ket{x_j} & \ket{A_j^1} & \cdots & \ket{A_j^{f(n_e)}} \\};
                \addvmargin{1mm}
            \end{tikzpicture}
        };
        \node at (8,0.6) {
            \begin{tikzpicture}[baseline=0,MyStyle/.style={draw, minimum width=2em, minimum height=1.8em, outer sep=0pt},every node/.style={font=\scriptsize}]
                \matrix (A) [matrix of math nodes, nodes={MyStyle, anchor=center}, column sep=-\pgflinewidth]
        {e_1 & e_2 & e_3 & e_4 & \cdots & \cdots & \cdots \\};
                \addvmargin{1mm}
            \end{tikzpicture}
        };
        \end{tikzpicture}
    \end{adjustbox}
    \caption{Quantum packet structure. The header carries quantum information for the quantum routing logic, such as source-destination quantum network addresses $\ket{x_i},\ket{x_j}$ and quantum superpositions $\ket{A_j}$ representing set of quantum nodes introduced in Sec.~\ref{sec:4.2}, while the payload carries entangled qubits (ebits) to be shared among network nodes.}
    \label{fig:02}
    \hrulefill
\end{figure}

The proposed quantum addressing scheme requires the definition of a \textit{quantum packet structure}, represented in Fig.~\ref{fig:02}. Differently from existing proposals \cite{DiAQiMil22,ThoKanKum-23}, our proposal consists of both a quantum header and a quantum payload, thus enabling a fully quantum-native processing. Specifically, the quantum packet is organized into:
\begin{itemize}
    \item a \textit{quantum header}, carrying quantum addresses\footnote{As detailed in the next two sections, a quantum address can be the univocal identifier of a network node, which by design is an orthogonal basis states. But it can also be a superposition of quantum states constituting the identifiers of a set of nodes -- as instance, the superposed address stored within the e-neighborhood entry of the routing table as shown in Fig.~\ref{fig:06}. As aforementioned, there may be also the need of sharing classical information, as instance, for classical signaling. Although quantum/classical coexistence is a key open problem, we envision that this can be achieved by multiplexing the quantum header/payload with classical header through quantum/classical multiplexing techniques  \cite{DiAQiMil22,ThoKanKum-23,ThoFeiChe-24}.}. This header enables the network nodes to interpret and forward entangled packets according to the quantum routing logic, based on the hierarchical address structure;
    \item and a \textit{quantum payload}, carrying entanglement as multiple ebits, intended for distribution to the destination node(s).
\end{itemize}
As aforementioned, this design allows the quantum packet to be processed in a fully quantum-native manner, without fallback to classical parsing mechanisms. This facilitates scalable quantum routing, as analyzed in the following section. Furthermore, since in this paper we focus on tier-2 nodes, we restrict our attention to packets carrying ESP-related information. Accordingly, the source and destination addresses $\{\ket{x_\ell}\}$ in Fig.~\ref{fig:02} should be interpreted as $\{\ket{v_\ell}\}$ and $\{ \ket{A_j^l} \}$ denote superposed quantum addresses representing set of nodes. 

As a final remark, although a comprehensive treatment is beyond the scope of this work and left for future investigation, the proposed quantum packet structure can be generalized to the multipartite entanglement case. In such a scenario, we envision the header carrying source address and a variable-length list of destination addresses, while the payload carrying a multipartite entangled state, which is intended for simultaneous distribution across the specified destination nodes. 

\section{Quantum-Native Routing}
\label{sec:3}
Here we design two different versions of the quantum-routing protocol tasked with proactively maintaining and utilizing the entanglement-activated overlay topology among the ESPs. Although these protocols differs in terms of overall control-plane complexity, they both leverage the quantum addressing scheme introduced in the previous section. 

Before delving into the technical details of the proposal, we refer the reader to Box~\ref{box:01}, which provides an outline of the overall routing logic for tier-1 nodes, which, as already described, delegate routing complexity to their serving ESPs.

\stepboxcounter{box:01}
\begin{figure*}[t]
\begin{tcolorbox}[colback=magenta!5!white,  
  colframe=magenta!80!black, 
  title=Box $1$: Outlook for Bottom Tier Nodes,
  sharp corners=south,
  boxrule=0.8pt,
  fonttitle=\bfseries,
  coltitle=black]
As aforementioned, by hybridizing classical hierarchical design principles with quantum features in the addressing scheme, the proposed architecture enables a \textit{tier-aware differentiation of routing information} stored locally at each node. Specifically, the structure of the quantum addresses, based on a shared prefix between tier-1 nodes and their serving ESP, allows tier-1 nodes to operate with minimal routing intelligence, effectively delegating forwarding decisions and complexity to tier-2 nodes. As a result, tier-1 nodes require only basic quantum-packet forwarding logic based on the generalization to the quantum domain of the classical prefix-matching mechanism. In contrast, ESPs maintain richer and more dynamic routing tables to manage inter-ESP overlay, activated by entanglement, as analyzed in Sec.~\ref{sec:3}. These tables are used to execute entanglement operations/manipulations, such as entanglement swapping, enabling end-to-end distribution of ebits across the network. This tiered routing abstraction significantly enhances scalability, by localizing complexity at tier-2 nodes. Indeed, it reduces both memory and processing overhead on tier-1 nodes, without compromising global routing capabilities.\\
More in detail, by leveraging the proposed quantum addressing scheme, end-to-end entanglement requests initiated by tier-1 nodes are first routed to their respective serving ESPs. In the ideal case, where the ESP overlay topology activated by the entanglement is fully connected, the serving ESP can directly perform entanglement swapping on the ebit(s) within the quantum payload, to reach the destination's serving ESP, thereby enabling efficient end-to-end entanglement distribution. \\
In more general and realistic scenarios, where the ESP overlay is not fully connected, the serving ESP must inspect its quantum routing table to assess whether pre-established entanglement exists with the destination's serving ESP. If such a ``link'' is available, the ESP can immediately perform entanglement swapping to enable end-to-end entanglement between the tier-1 nodes. Otherwise, the ESP must identify an appropriate next-hop ESP (in the generalized sense analyzed above), selected according to the adopted routing metric, that brings the quantum packet ``closer'' (again, in the generalized sense) to the destination. Thus, the ebit(s) in the quantum packet payload is eventually forwarded through an entangled path, i.e., through a sequence of intermediate ESPs, until an ESP already sharing entanglement with the destination ESP is found. This strategy, ruled by the quantum addressing scheme and the selected routing metric, is detailed in Sec.~\ref{sec:3}.\\
While several alternative design choices for tier-1 nodes are possible, including the option to handle their requests only with classical communications\footnote{In such a case the proposed quantum packet structure would apply only to ESPs.}, this paper focuses on the core routing functionality of ESPs, which ensures end-to-end entanglement distribution across the network. 
Accordingly, the exploration of alternative tier-1 design choices lies beyond the scope of this work. 
\end{tcolorbox}
\end{figure*}

\subsection{Preliminaries}
\label{sec:3.1}
As pioneered in \cite{CacIllCal-23} and reflected in the previous sections, the \textit{goal} of a quantum routing protocol for ESPs fundamentally diverges from classical routing paradigm. In contrast to the classical objective of determining physical routes toward destinations for packet forwarding, a native-quantum routing shifts the goal to:
\begin{quote}
    \textit{proactively maintaining the overlay topology activated by the entanglement and ensuring end-to-end entanglement distribution, by managing and tracking the entangled resources} shared within the tier-2 network.
\end{quote}

Achieving this goal requires a paradigmatic departure in the structure and semantics of routing tables.  Specifically, routing entries no longer store next-hop interfaces toward destination addresses. Rather, they list the locally-available entangled qubits\footnote{As instance, as pointers to the ``addresses'' of the communication qubits \cite{rfc9340} storing the entangled state.} at each node, together with the identities\footnote{Indeed, any quantum communication protocol requires a tight cooperation between the network nodes storing the entangled qubits for being able to exploit the quantum correlation provided by entanglement, and thus nodes must be aware of each other identities.} of the ESPs sharing this entanglement.

This shift does not imply that physical topology knowledge is irrelevant. On the contrary, such knowledge remains essential for the generation and distribution of link-level entanglement, needed to replenish or restore depleted entangled resources. In this perspective, by leveraging the dual-addressing framework described above in which classical addresses are augmented with quantum addresses, each ESP maintains two distinct routing tables:
\begin{itemize}
    \item  a classical routing table, used to manage physical topological information related to the classical communication infrastructure and populated using classical routing protocols;
    \item  a quantum routing table, used to manage the entangled overlay topology information  and populated by the EDC.
\end{itemize}

The focus of the remaining part of the manuscript is on the design of such quantum routing tables and on their central role in enabling scalable, quantum-native forwarding decisions within the tier-2 network. To this aim, we first collect some definitions used in the following.

\begin{defin}[\textbf{Link Entanglement Metric}]
    \label{def:02}
    The ``cost'' associated with generating and distributing \textit{link entanglement} over a quantum link $(i,j) \in V_2 \times V_2$ is modeled by a general \textit{entangling-cost metric} $w(i,j)$. This metric is equipped with two operations \cite{Sob-05}:
    \begin{itemize}
        \item[i)] an order relation $<$,  allowing pairwise comparison of link entanglement costs;
        \item[ii)] a binary operation $\oplus$, enabling the composition of entanglement costs over multiple links.
    \end{itemize}
    In particular, the order operation $<$ allows us to express relative entangling difficulty across links. For instance:
    \begin{equation}
        \label{eq:09}
        w(i,k) < w(j,k),
    \end{equation}
    implies that establishing link entanglement between ESP $v_i$ and $v_k$ is less costly than between $v_j$ and $v_k$.
    The binary operation $\oplus$ extends the cost function from individual links to multi-hop entanglement paths. For instance:
    \begin{equation}
        \label{eq:10}
        w\big( (i,k) \oplus (k,j) \big)
    \end{equation}
    denotes the cost of distributing end-to-end entanglement between remote ESPs $v_i$ and $v_j$, via an intermediate ESP $v_k$.
\end{defin}

This abstraction allows us to support arbitrary entanglement cost models, which in turn reflect the chosen quantum routing metric \cite{Cal-17}. In other words, with this abstraction, \textit{we decouple the routing logic from any specific cost formulation}. Hence, the framework accommodates a broad class of quantum metrics, ranging from fidelity degradation to quantum memory usage or decoherence effects. Consequently, the entangling cost $w(\cdot,\cdot)$ encapsulates the link/route ``quality'' under the selected model, allowing the routing protocol to adapt to the specific characteristics and constraints of the underlying quantum network.

\begin{remark}
    For notational and conceptual simplicity, we refer to routes with the lowest entangling cost as \textit{shortest paths}. This is in line with the conventional terminology of classical routing, originally developed under the assumption that the ``quality'' of a communication path is determined by its hop count, with shorter paths being preferable. 
\end{remark}

We assume that $w(\cdot,\cdot)$ satisfies the axiomatic properties of a metric, i.e.:
\begin{align}
    \label{eq:11}
    & \text{definiteness: } & w(i,i) = 0 \Longleftrightarrow v_i = v_j \\\
    \label{eq:12}
    & \text{non-negativity: } & w(i,j) \geq 0 \\
    \label{eq:13}
    & \text{symmetry: } & w(i,j) = w(j,i) \\
    \label{eq:14}
    & \text{triangle inequality: } & w(i,j) \leq w\big( (i,k) \oplus (k,j) \big), 
\end{align}
plus an additional property to ensure the metric being isotone, namely, being both left-isotone:
\begin{align}
    \label{eq:15}
    w(j,l) < w(j,k) \Longrightarrow w\big( (i,j) \oplus (j,l) \big) < w\big( (i,j) \oplus (j,k) \big),
\end{align}
and right-isotone:
\begin{align}
    \label{eq:16}
    w(l,j) < w(k,j) \Longrightarrow w\big( (l,j) \oplus (j,i) \big) < w\big( (k,j) \oplus (j,i) \big).   
\end{align}

\begin{remark}
    Isotonicity ensures that the relative ordering of entangling costs between two quantum links (or paths), sharing a common origin or destination, is preserved when both are extended by the same quantum link. Meanwhile, triangle inequality, also referred to as monotonicity in \cite{Sob-05}, implies that the entangling cost cannot decrease, when it is extended by a new quantum link. These properties are not only rationale for capturing the constraints\footnote{As an example, the triangle inequality correctly models the preference for direct link entanglement over entanglement swapping at an intermediate node.} of entanglement distribution, but they are required for guaranteeing the convergence \cite{Cal-17} of quantum routing protocols to optimal entanglement paths, without resorting to exhaustive enumeration of all possible routes \cite{Sob-05}.
\end{remark}

\begin{defin}[\textbf{End-to-end Entangling Metric}]
    \label{def:03}
    Let us consider two pair of non-adjacent ESPs, say $v_i$ and $v_j$ and let $R$ denote a set of ESPs forming a valid swapping path from $v_i$ to $v_j$. $w_R(i,j)$ denotes the accumulated entangling cost along that specific path, using the $\oplus$ operator. Clearly, end-to-end entanglement between $v_i$ and $v_j$ can be achieved through multiple possible routes, and an effective routing protocol should select the route with the lowest entangling cost. Specifically, swapping should occur at the most favorable, accordingly to the selected metric, sequence of intermediate ESPs acting as quantum repeaters, i.e.:
    \begin{equation}
        \label{eq:17}
        w(i,j) = \min_{R} \{ w_R(i,j) \}.
    \end{equation}
\end{defin}

Stemming from Def.~\ref{def:03}, we are now ready to formally define the \textit{entangling stretch} introduced by a quantum routing protocol. Broadly speaking, the purpose of a quantum routing protocol is to discover and select, among the available ESPs, a suitable sequence of intermediate nodes that enables two remote ESPs to establish shared entanglement through entanglement swapping. This selection is driven by a predefined cost metric, which models the quality or difficulty of establishing entanglement along a link or path.

Let $\mathcal{QR}$ denote an arbitrary quantum routing protocol. And let us define with $R = \mathcal{QR}(i,j)$ the set of ESPs selected as repeaters by $\mathcal{QR}$ to establish end-to-end entanglement between the remote (in the overlay topology) ESPs $v_i$ and $v_j$.

\begin{defin}[\textbf{Entangling Stretch}]
    \label{def:04}
    The entangling stretch induced by the quantum routing protocol $\mathcal{QR}$ between ESPs $v_i$ and $v_j$ is defined\footnote{For simplicity, we omit the explicit dependence of $\mathscr{ES}(i,j)$ on $\mathcal{QR}$, i.e., $\mathscr{ES}(i,j) \eqdef \mathscr{ES}_{\mathcal{QR}}(i,j)$.} as:
    \begin{equation}
        \label{eq:18}
        \mathscr{ES}(i,j) = \frac{w_{R}(i,j)}{w(i,j)}, \quad \text{with } R = \mathcal{QR}(i,j),
    \end{equation}
    where $w_R(i,j)$ denotes the entangling cost incurred by the path selected by the routing protocol $\mathcal{QR}$, and $w(i,j)$ represents the optimal entangling cost in \eqref{eq:17}.
    Accordingly, the \textit{overall entangling stretch} $\mathscr{ES}$ of the quantum routing protocol $\mathcal{QR}$ is defined as the worst-case stretch over all ESP pairs in the network:
    \begin{equation}
        \label{eq:19}
        \mathscr{ES} = \max_{v_i,v_j \in V_2} \left\{ \mathscr{ES}(i,j) \right\}.
    \end{equation}
\end{defin}

From Eq.~\eqref{eq:19}, it follows that a quantum routing protocol is optimal if and only if its entangling stretch satisfies $\mathscr{ES} = 1$, since it always selects the lowest-cost entangled path for any ESP pair. Conversely, an entangling stretch strictly greater than $1$ indicates that the protocol may route entanglement through suboptimal entangled paths, for at least some pairs of ESPs.

\begin{remark}
    We emphasize that the term \emph{stretch}, used in Def.~\ref{def:04}, stems from classical routing terminology. In classical networks, the \emph{path stretch} of a routing scheme is typically defined as the ratio between the actual path length followed by a packet (usually measured in number of hops) and the shortest possible path length\footnote{In this context, the research area in classical networks named as \textit{compact routing} aims at designing scalable routing schemes able to guarantee a path stretch upper-bounded by a \textit{constant factor $c$} independent from the network size $n$, such as $c=5$ or even $c=3$ as in \cite{ThoZwi-01}. In this research area, both the table-scaling and the path-stretch bounds are derived with a worst-case analysis, i.e., the routing table of each node scales sub-linearly in $n$ and the path stretch is at most $c$ among all the source-destination paths}\cite{KriClaFal-07}. Here we adopt this classical terminology, by reinterpreting the concept: the stretch is defined as the ratio between the entanglement cost incurred by the routing protocol, based on the selected sequence of repeaters, and the minimum possible entanglement cost under optimal swapping sequence. This generalization preserves the intuition behind the term ``stretch'', while capturing the fundamentally different nature of cost in quantum networks, reflecting entanglement quality/complexity.
\end{remark}

\subsection{Design Principles}
\label{sec:3.2}

\begin{design}[\textbf{Compact}]
    \label{Des:02}
    Our goal is to design a quantum routing protocol with \textit{sublinear quantum memory requirements}. This is achieved by defining: i) routing tables that, for each ESP, have at most\footnote{\label{foot:02} In this manuscript, we adopt the computer science notation for classifying routing protocols time/space complexity. Accordingly, $\mathcal{O}(\cdot)$ (namely, \text{big O}) denotes the asymptotical growth rate of the number of communication qubits stored at each node as the network size grows, while $\tilde{\mathcal{O}}(\cdot)$ (namely, \text{soft O}) denotes the asymptotical growth rate when logarithmic factors are ignored.} $\tilde{\mathcal{O}}(\sqrt{n_e})$ entries, and ii) \textit{logarithmic} quantum network addresses, i.e., addresses that scales in size as $\mathcal{O}(\log {n})$, with $n$ denoting the number of ESPs within the quantum network.
\end{design}

The design principle~\ref{Des:02} is very reasonable, since maintaining entanglement between every pair of ESPs would determine prohibitive memory requirements at each ESP, as well as significant network overhead in terms of both quantum and classical resources for entanglement generation and distribution. Consequently, we adopt the strategy whereby each ESP shares entanglement with only a subset of the other ESPs. And the size of this subset scales sublinear with the total number of ESPs, i.e., $\sqrt{n_e} \log n_e$ or equivalently\footref{foot:02} $\tilde{\mathcal{O}}(\sqrt{n_e})$.

Clearly, the choice of which ESP belongs to such a fraction crucially dictates the overall quantum routing performance. This aspect is thoroughly analyzed in Secs.~\ref{sec:3.3}-Sec.~\ref{sec:3.4}, where we introduce and compare two distinct design strategies, offering different trade-offs in terms of complexity and routing efficiency.

\begin{table}
    \renewcommand{\arraystretch}{1.4}
    \centering
    \begin{tabular}{l c c c c}
        \multicolumn{5}{c}{\textsc{entangling stretch upper bound}}\\
        \hline
        \hline
        \textbf{routing scheme} & \multicolumn{3}{c}{\textbf{entangling cost metric}} & \textbf{section}\\
        \cline{2-4}
        & arbitrary & additive & $\min$ \\
        \hline
        \hline
        Partial-Anchor & $\displaystyle \frac{w\big( \oplus_5 (i,j) \big) }{w(i,j)}$ & $5$ & $1$ & Sec.~\ref{sec:3.3} \\
        \hline
        Full-Anchor & $\displaystyle \frac{w\big( \oplus_3 (i,j) \big) }{w(i,j)}$ & $3$ & $1$  & Sec.~\ref{sec:3.4}\\
        \hline
        \hline
    \end{tabular}
    \caption{Entangling stretch vs. quantum routing scheme. With a ``Partial-Anchor'' scheme, only a subset of ESPs is responsible for proactively creating long-range (i.e., high-cost) artificial links as shown in Fig.~\ref{fig:05}, yet at the price of a slight increased entangling stretch.}
    \label{tab:03}
    \hrulefill
\end{table}

\begin{design}[\textbf{Constant Stretch}]
    \label{Des:03}
    Our objective is to design a quantum routing protocol that guarantees an \textit{overall entangling stretch} $\mathscr{ES}$ upper-bounded by a constant factor for an arbitrary isotonic entangling cost metric: 
    \begin{equation}
        \label{eq:20}
        \mathscr{ES} \leq c \eqdef \frac{w_{\tilde{R}}(i,j)}{w(i,j)},
    \end{equation}
    where $v_i$ and $v_j$ denote the ESPs exhibiting the worst-case entangling stretch, i.e., the pair of nodes satisfying  \eqref{eq:19}, and $w_{\tilde{R}}(i,j)$ denotes an upper bound on the cost of entangling $v_i$ and $v_j$  through the entangled path discovered by the protocol. 
    The exact expression of the constant factor depends on the specific design choices.
\end{design}

\begin{remark}
    In Secs.~\ref{sec:3.3} and \ref{sec:3.4}, we show that our protocol is able to assure a constant stretch of 5 and 3, respectively, i.e.:
    \begin{equation}
        \label{eq:21}
        \mathscr{ES} \leq c = 3 \vee 5.
    \end{equation}
    These results are obtained under the assumption that the entangling cost metric is \textit{additive}, namely, when the binary operation $\oplus$ in Def.~\ref{eq:02} denotes the standard addition: $w\big( (i,k) \oplus (k,j) \big) \eqdef w(i,k) + w(k,j)$. It is worthwhile to note that routing protocol metrics (classical or quantum) are mainly additive, with delay and the quantum version of the hop count as representative examples. Other popular metrics could be either multiplicative, e.g., packet loss, or concave metrics, e.g, bandwidth. As for the former class, multiplicative metrics can be straightforwardly converted into additive metrics via logarithmic scaling. Differently, for the latter class -- namely, whenever the binary operation $\oplus$ denotes the minimum operator \cite{RieSchDom-07} and $w\big( (i,k) \oplus (k,j) \big) \eqdef \min \{w(i,k), w(k,j)\}$ -- the constant factor is unitary:
    \begin{equation}
        \label{eq:22}
        \mathscr{ES} = 1.
    \end{equation}
    From the above, it is evident that restricting the attention on additive entangling-cost metric $w(\cdot,\cdot)$ is not a limitation. On the contrary, it represents a conservative design choice: the concave case trivially achieves optimality, while additive metrics pose a more realistic and challenging setting for protocol design.
\end{remark}

Although the next design principle has been already introduced and justified in the previous section, here we report it once again for the sake of completeness.

\begin{design}[\textbf{Quantum Addressing}]
    \label{Des:04}
    Our objective is to design a quantum routing protocol that takes advantage of quantum principles and phenomena via quantum addressing.
\end{design} 

Summarizing, regardless of the particulars of the adopted entangling cost metric $w(\cdot,\cdot)$, which determines the final form of \eqref{eq:19}, our design principles enforce a quantum routing protocol characterized by the following key features and summarized in Table~\ref{tab:04}:
\begin{itemize}
    \item[i)] \textit{Sublinear size of the routing tables}: each ESP maintains entanglement with an ESP set of size $\tilde{\mathcal{O}}\left( \sqrt{n_e} \right)$, where $n_e$ is the total number of ESPs. This implies a sublinear quantum memory overhead.
    \item[ii)] \textit{Logarithmic quantum address length}: the quantum address length scales as $\mathcal{O}\left( \log n \right)$, with $n$ denoting the total number of nodes in the network.
    \item[iii)] \textit{Constant entangling stretch}: the quantum communication overhead, measured in terms of entangling stretch, and quantifying the entanglement ``wasted'' for not having a fully-connected overlay mesh among the ESPs, is upper bounded by a constant, independent of the network size.
\end{itemize}

\begin{table*}[t]
    \centering
    \renewcommand{\arraystretch}{1.4}
    \begin{tabular}[\columnwidth]{l|l|p{0.65\linewidth}}
        \hline
        \hline
        \textbf{metric} & \textbf{scaling} & \textbf{description} \\
        \hline
        \hline
        quantum memory overhead & $\tilde{\mathcal{O}}(\sqrt{n_e})$ & size of the ESP Routing Table defining the number of entangled links maintained by each ESP node, with $n_e$ denoting the total number of ESPs \\
        \hline
        quantum address length & $\mathcal{O}(\log n)$ & number of qubits required to encode a quantum address, with $n$ denoting the total number of network nodes\\
        \hline
        entangling stretch & $\mathscr{ES} \leq c$ & maximum ratio between the actual and optimal entangling cost bounded by a constant $c$, independent of network size\\
        \hline
        \hline
    \end{tabular}
    \caption{Key performance indicators of the proposed quantum-native routing protocol.}
    \label{tab:04}
    \hrulefill
\end{table*}

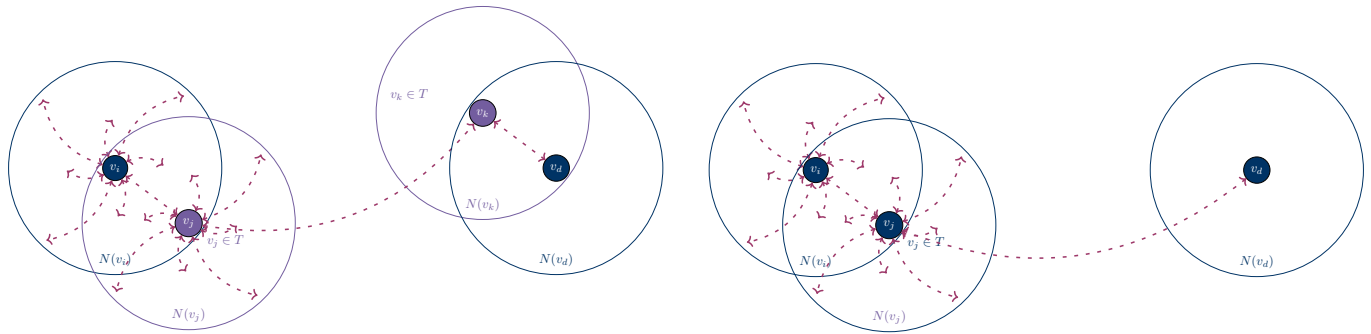
\begin{figure*}
	\centering
	\begin{minipage}[c]{0.48\textwidth}
		\centering
        \begin{adjustbox}{width=\columnwidth}
            \begin{tikzpicture}[->]
                \node [circle, draw, fill=myBlue, text=white] (i) at (0,0) {$v_{i}$};
                \draw[fill=none,color=myBlue] (0,0) circle (2.9) node [myBlue,yshift=-2.5cm] {$N(v_i)$};
                \node (i0) at (-2,-2) {};
                \draw[very thick,loosely dashed,<->,bend left,myPurple] (i) edge (i0);
                \node (i1) at (-1.44, 0) {};
                \draw[very thick,loosely dashed,<->,bend left,myPurple] (i) edge (i1);
                \node (i2) at (-2, 2) {};
                \draw[very thick,loosely dashed,<->,bend left,myPurple] (i) edge (i2);
                \node (i3) at (0, +1.44) {};
                \draw[very thick,loosely dashed,<->,bend left,myPurple] (i) edge (i3);
                \node (i4) at (2, 2) {};
                \draw[very thick,loosely dashed,<->,bend left,myPurple] (i) edge (i4);
                \node (i5) at (+1.44, 0) {};
                \draw[very thick,loosely dashed,<->,bend left,myPurple] (i) edge (i5);
                \node (i7) at (0, -1.44) {};
                \draw[very thick,loosely dashed,<->,bend left,myPurple] (i) edge (i7);
                
                \node [circle, draw, fill=myTrackingSet, text=white] (j) at (2,-1.5) {$v_{j}$};
                \node[myTrackingSet] at (3,-2) {$v_j \in T$};
                \draw[fill=none,color=myTrackingSet] (2,-1.5) circle (2.9) node [myTrackingSet,yshift=-2.5cm] {$N(v_j)$};
                \draw[very thick,loosely dashed,<->,myPurple] (i) edge (j);
                \node (j0) at (0,-3.5) {};
                \draw[very thick,loosely dashed,<->,bend right,myPurple] (j) edge (j0);
                \node (j1) at (0.66, -1.5) {};
                \draw[very thick,loosely dashed,<->,bend right,myPurple] (j) edge (j1);
                \node (j3) at (2, -0.06) {};
                \draw[very thick,loosely dashed,<->,bend right,myPurple] (j) edge (j3);
                \node (j4) at (4, 0.5) {};
                \draw[very thick,loosely dashed,<->,bend right,myPurple] (j) edge (j4);
                \node (j5) at (3.44, -1.5) {};
                \draw[very thick,loosely dashed,<->,bend right,myPurple] (j) edge (j5);
                \node (j6) at (4, -3.5) {};
                \draw[very thick,loosely dashed,<->,bend right,myPurple] (j) edge (j6);
                \node (j7) at (2, -2.94) {};
                \draw[very thick,loosely dashed,<->,bend right,myPurple] (j) edge (j7);

                \node [circle, draw, fill=myTrackingSet, text=white] (k) at (10,1.5) {$v_{k}$};
                \node[myTrackingSet] at (8,2) {$v_k \in T$};
                \draw[fill=none,color=myTrackingSet]  (10,1.5) circle (2.9) node [myTrackingSet,yshift=-2.5cm] {$N(v_k)$};
                \draw[very thick,loosely dashed,<->,bend right, myPurple] (j) edge (k);
                
                \node [circle, draw, fill=myBlue, text=white] (d) at (12,0) {$v_{d}$};
                \draw[fill=none,color=myBlue] (12,0) circle (2.9) node [myBlue,yshift=-2.5cm] {$N(v_d)$};
                \draw[very thick,loosely dashed,<->,myPurple] (k) edge (d);
            \end{tikzpicture}
        \end{adjustbox}
    	\subcaption{Partial-Anchor Scheme: only the fraction of ESPs in $T$ (depicted in purple) is responsible for proactively creating high-cost artificial links each others, whereas all the ESPs maintain low-cost artificial links with their e-neighborhood $N(\cdot)$.}
		\label{fig:05a}
	\end{minipage}
	\hspace{0.02\textwidth}
	\begin{minipage}[c]{0.48\textwidth}
		\centering
        \begin{adjustbox}{width=\columnwidth}
            \begin{tikzpicture}[->]
                \node [circle, draw, fill=myBlue, text=white] (i) at (0,0) {$v_{i}$};
                \draw[fill=none,color=myBlue] (0,0) circle (2.9) node [myBlue,yshift=-2.5cm] {$N(v_i)$};
                \node (i0) at (-2,-2) {};
                \draw[very thick,loosely dashed,<->,bend left,myPurple] (i) edge (i0);
                \node (i1) at (-1.44, 0) {};
                \draw[very thick,loosely dashed,<->,bend left,myPurple] (i) edge (i1);
                \node (i2) at (-2, 2) {};
                \draw[very thick,loosely dashed,<->,bend left,myPurple] (i) edge (i2);
                \node (i3) at (0, +1.44) {};
                \draw[very thick,loosely dashed,<->,bend left,myPurple] (i) edge (i3);
                \node (i4) at (2, 2) {};
                \draw[very thick,loosely dashed,<->,bend left,myPurple] (i) edge (i4);
                \node (i5) at (+1.44, 0) {};
                \draw[very thick,loosely dashed,<->,bend left,myPurple] (i) edge (i5);
                \node (i7) at (0, -1.44) {};
                \draw[very thick,loosely dashed,<->,bend left,myPurple] (i) edge (i7);
                
                \node [circle, draw, fill=myBlue, text=white] (j) at (2,-1.5) {$v_{j}$};
                \node[myBlue] at (3,-2) {$v_j \in T$};
                \draw[fill=none,color=myBlue] (2,-1.5) circle (2.9) node [myTrackingSet,yshift=-2.5cm] {$N(v_j)$};
                \draw[very thick,loosely dashed,<->,myPurple] (i) edge (j);
                \node (j0) at (0,-3.5) {};
                \draw[very thick,loosely dashed,<->,bend right,myPurple] (j) edge (j0);
                \node (j1) at (0.66, -1.5) {};
                \draw[very thick,loosely dashed,<->,bend right,myPurple] (j) edge (j1);
                \node (j3) at (2, -0.06) {};
                \draw[very thick,loosely dashed,<->,bend right,myPurple] (j) edge (j3);
                \node (j4) at (4, 0.5) {};
                \draw[very thick,loosely dashed,<->,bend right,myPurple] (j) edge (j4);
                \node (j5) at (3.44, -1.5) {};
                \draw[very thick,loosely dashed,<->,bend right,myPurple] (j) edge (j5);
                \node (j6) at (4, -3.5) {};
                \draw[very thick,loosely dashed,<->,bend right,myPurple] (j) edge (j6);
                \node (j7) at (2, -2.94) {};
                \draw[very thick,loosely dashed,<->,bend right,myPurple] (j) edge (j7);

                \node (k) at (10,4.4) {};

                \node [circle, draw, fill=myBlue, text=white] (d) at (12,0) {$v_{d}$};
                \draw[fill=none,color=myBlue] (12,0) circle (2.9) node [myBlue,yshift=-2.5cm] {$N(v_d)$};
                \draw[very thick,loosely dashed,<->,bend right,myPurple] (j) edge (d);
            \end{tikzpicture}
        \end{adjustbox}
    	\subcaption{Full-Anchor Scheme: each ESP is responsible for proactively creating both low- and high-cost artificial links, the formers with their e-neighborhood $N(\cdot)$ and the latters with a specific subset of nodes.}
		\label{fig:05b}
	\end{minipage}
    \caption{Schematic view of the two schemes underlying our  compact quantum routing protocol: \textit{Partial-Anchor} (presented in Sec.~\ref{sec:3.3}) vs \textit{Full-Anchor} (presented in Sec.~\ref{sec:3.4}). Artificial links are denoted with red arrows, whereas circles around ESPs denote their e-neighborhood as defined in Def.~\ref{def:05}.}
    \label{fig:05}
    \hrulefill
\end{figure*}

\subsection{Partial-Anchor Scheme}
\label{sec:3.3}

Here, as aforementioned, we design the first version of the quantum-routing protocol, in which a small subset of ESPs, referred to as \emph{anchor ESPs}, is responsible for proactively establishing and maintaining long-range (i.e., high-cost) entangled links. The remaining ESPs, constituting the majority, are instead in charge of proactively creating and maintaining short-range (i.e., low-cost) artificial links.

As proved in the following, this design yields an overlay topology, where any ESP is at most 3-hop-entanglement away from any other ESP (see Remark~\ref{rem:06}). This worst-case guarantee stems from the strategic connectivity established through the anchor ESPs, which serve as entanglement hubs within the overlay.

Conversely, when we relax the constraint of (very) limited number of anchor ESPs, as explored in Sec.~\ref{sec:3.4}, the overlay topology is restructured to have every ESP 2-hop-entanglement distant from any other ESP (see Remark~\ref{rem:09}). Thus, we reduce the communication cost in terms of entangling stretch, at the price of increasing the maintenance complexity of the overlay. Remarkably, this enhancement in routing performance is achieved without affecting the asymptotic scaling of the routing table size.

The aforementioned trade-off between routing performance and control complexity, summarized in Table~\ref{tab:03}, gives rise to two distinct protocol variants: the first, governed by the following last-but-not-least design principle, is referred to as \textit{partial-anchor scheme}; the second, discussed in the next section and formalized by Design Principle~\ref{Des:06}, is referred to as the \textit{full-anchor scheme}.

\setcounter{design}{4}
\begin{subdesign}[\textbf{Partial-Anchor Scheme}]
    \label{Des:05}
    Our objective is to design a quantum routing protocol where most of the nodes proactively creates and maintains short-range (i.e., low-cost) artificial links, and only a fraction of nodes is responsible for proactively creating and maintaining long-range (i.e., high-cost) artificial links.
\end{subdesign} 

Accordingly, we begin by defining two sets of nodes: the \textit{e-neighborhood} and the \textit{anchor-nodes}.

\begin{defin}[\textbf{e-neihborhood}]
    \label{def:05}
    The e-neighborhood\footnote{We omit the explicit dependence on $k$ for notational simplicity.} $N(v_i)$ of the arbitrary ESP $v_i \in V_2$ denotes the set of the $k = |N(v_i)|$ nodes closest
    to $v_i$, according to the entangling-cost metric $w(\cdot,\cdot)$. The nodes in $N(v_i)$ are referred to as the \textit{e-neighbors} of $v_i$.
\end{defin}
Intuitively, the e-neighborhood of an ESP captures also localized information about the underlying physical topology, as inferred through the entangling-cost $w(\cdot,\cdot)$. In this sense, the concept of e-neighborhood inherently reflects the low-cost regime associated with the proactive creation and maintenance of entangled links. Conversely, only the nodes in the following set are requested to proactively create and maintain high-cost entangled links.

\begin{table*}[t]
    \centering
    \renewcommand{\arraystretch}{1.3}
    \begin{tabular}{p{0.2\linewidth}|p{0.3\linewidth}|p{0.3\linewidth}}
        \hline
        \hline
        \textbf{feature} & \textbf{e-neighborhood} & \textbf{anchor} \\
        \hline
        \hline
        \textbf{connectivity scope} & local & global (long-range) \\
        \hline
        \textbf{entanglement} & toward the $k$-closest ESPs according to entangling cost $w(\cdot,\cdot)$ & toward the other $|T|-1$ anchors \\
        \hline
        \textbf{entanglement cost} & low & high \\
        \hline
        \textbf{routing role} & capturing local physical topology structure and ensuring short-entanglement hop redundancy & enabling global connectivity across distant ESPs \\
        \hline
        \textbf{overlay topology impact} & supporting dense local clustering & guaranteeing that any pair of ESP is within at most 3-entanglement-hops \\
        \hline
        \textbf{maintenance complexity} & low & high \\
        \hline
        \textbf{design parameter} & number of e-neighbors $k = |N(v_i)|$ & number of anchors $|T|$ \\
        \hline
        \textbf{scalability/performance impact} & sublinear routing tables & constant entangling stretch \\
        \hline
        \hline
    \end{tabular}
    \caption{Comparison between e-neighborhood $N(\cdot)$ and anchor set $T$, highlighting their respective contributions to local and global entanglement reachability in the overlay topology.}
    \label{tab:05}
    \hrulefill
\end{table*}

\begin{defin}[\textbf{Anchor Set}]
    \label{def:06}
    A subset $T \subseteq V_2$ of ESPs is defined as \textit{anchor set}, and its elements are equivalently referred to as \textit{anchor ESPs} or simply \textit{anchors}.
\end{defin}

From the above, it is evident that the two key parameters influencing the overall routing behavior are: i) the size of each ESP e-neighborhood, denoted by $k$, and ii) the cardinality of the anchor-node set $|T|$, as summarized in Tab.~\ref{tab:05}. Building on this observation, we now propose a constructive method to set these parameters to induce the compact routing Property~\ref{proper:01}, which is subsequently exploited by the routing protocol. 

Specifically, we leverage a result from classical combinatorial set theory concerning the covering set problem, originally introduced in \cite{Lov-75}, but in the formulation given in \cite{AweBarLin-89} as follows. Let $P$ be a finite set of $\ell$ elements, and let $\mathcal{A} = \{A_1, \ldots, A_h\}$ be a collection of subsets of $P$, each with cardinality $|A_i| = s$. A set $D \subseteq P$ is said to cover $\mathcal{A}$ if:
\begin{equation}
    \label{eq:23}
    \forall A_i \in \mathcal{A}, \quad A_i \cap D \neq \emptyset.
\end{equation}
By constructing the set $D$ using a randomized cover algorithm, then with high probability $1 - \mathcal{O}(h^{1-m})$, for any constant $m > 1$, $D$ covers $\mathcal{A}$ and its size is upper bounded as:
\begin{equation}
    \label{eq:24}
    |D| = \mathcal{O}\left( \frac{\ell \log h}{s} \right).
\end{equation}

In our context, we map this result to the quantum routing problem, by interpreting $P$ as the set of ESPs, and the collection $\mathcal{A}$ as the set of e-neighborhoods, each with size $k = |N(v_j)|$. The goal is to construct the anchor set $T$, such that each e-neighborhood $N(v_j)$ contains at least one anchor ESP. Thus, in this analogy, $T$ plays the role of the covering set $D$. By applying the result from \cite{AweBarLin-89} and by neglecting lower-order terms, we ensure that if the e-neighborhood size is set to:
\begin{equation}
    \label{eq:25}
     k = (1 + m) \sqrt{n_e} \log n_e,
\end{equation}
then there exists a cover set $T$ of size:
\begin{equation}
    \label{eq:26}
    |T| =  \sqrt{n_e},
\end{equation}
that covers all the e-neighborhoods with high probability. In other words, with this parametrization, we are able to ensure that the anchor set $T$ serves as a dominating set over the graph induced by e-neighborhoods, thereby enabling the following crucial Property~\ref{proper:01}, while maintaining both sublinear memory complexity and constant stretch.

\begin{proper}
    \label{proper:01}
    Any ESP has in its e-neighborhood at least one anchor with high probability (w.h.p.), i.e.:
    \begin{equation}
        \label{eq:27}
        P \big( \not\exists \, v_j \in T : v_j \in N(v_i) \big) = \mathcal{O}(n_e^{1-m}),
        \quad \forall \, v_i \in V_e.
    \end{equation}
\end{proper}

\begin{remark}
    \label{rem:05}
    As aforementioned, the anchor set $T$ can be constructed by using the randomized algorithm proposed in \cite{AweBarLin-89}, which does not require any a-priori knowledge nor any particular property. And the probability of violating Property~\ref{proper:01} can be made arbitrarily small, by appropriately tuning the constant $m$. In practical scenarios, when the anchors are equipped with enhanced quantum hardware capabilities\footnote{An assumption arguably reasonable given their role in proactively maintaining long-range entangled links.}, it may be preferable to deterministically construct the anchor set. This can be accomplished either by de-randomizing the aforementioned algorithm using techniques such as those proposed in~\cite{AbrGavMal-08}, or by adopting a classical greedy approach as originally introduced in~\cite{Lov-75}.
    In the latter case, a tighter bound on the anchor set size can be achieved, such as $|D| \leq \frac{\ell (1 + \log h)}{s}$, yet at the price of some global network knowledge. Indeed, more efficient covering algorithms can be devised, by exploiting particular properties of the overlay topology. However, the investigation of optimized techniques for the anchor-set construction lies beyond the scope of this paper. Our goal here is to demonstrate that Property~\ref{proper:01} can be enforced with high probability through careful parameterization, and that both randomized and deterministic construction methodologies exist to support this design objective.
\end{remark}

With the above in mind, the key design choice for the routing protocol is the following: 
\begin{quote}
    \textit{each anchor node $v_j \in T$ proactively establishes and maintains entanglement with any other anchor node in $T$,}
\end{quote}
via the optimal end-to-end entangling metric in Def.~\ref{def:03}. Accordingly, $v_i,v_j \in T$ are connected by an artificial link within the artificial topology, with the lowest entangling stretch, i.e., $\mathscr{ES}(i,j) = 1$. Additionally:
\begin{quote}
    \textit{each ESP $v_i$ proactively establishes and maintains entanglement with any ESP within its e-neighborhood $N(v_i)$},
\end{quote}
again, by exploiting the optimal entangling metric in Def.~\ref{def:03}. Such artificial links between $v_i$ and any $v_j \in N(v_i)$ likewise exhibit the minimum entangling stretch, i.e., $\mathscr{ES}(i,k) = 1$. This design is depicted in Fig.~\ref{fig:05a}, where non-anchor ESPs, i.e., nodes in $V_2 \setminus T$, and ESPs in $T$ are represented in blue and purple, respectively, and artificial links are denoted with red arrows.

\begin{figure*}
    \setlength{\tabcolsep}{4pt} 
    \def\arraystretch{1.1} 
    \centering
    \begin{NiceTabular}{c c c c c c}
        \Block{1-6}{\textsc{Quantum Routing Table at ESP} $v_i$} \\
        \hline
        \hline
        ebits & e-hop & e-neghborhood & Anchor \\
        \hline
        \hline
        \begin{tikzpicture}[baseline=0,MyStyle/.style={draw, minimum width=1.5em, minimum height=1.5em, outer sep=0pt}]
            \matrix (A) [matrix of math nodes, nodes={MyStyle, anchor=center}, column sep=-\pgflinewidth]
        {e_1 & e_2 & \cdots & \cdots \\};
            \addvmargin{1mm}
        \end{tikzpicture} & $\ldots$ & $\ldots$ & $\ldots$ & \Block{3-2}{for any $v_i \in V_e$:\\$\mathcal{O}(\sqrt{n_e} \log n_e)$ entries\\one for each $v_j : v_i \in N(v_j)$}\\
        \cline{1-4}
        \begin{tikzpicture}[baseline=0,MyStyle/.style={draw, minimum width=1.5em, minimum height=1.5em, outer sep=0pt}]
            \matrix (A) [matrix of math nodes, nodes={MyStyle, anchor=center}, column sep=-\pgflinewidth]
        {e_1 & e_2 & \cdots & \cdots \\};
            \addvmargin{1mm}
        \end{tikzpicture}
            & $\ket{v_j}$ & $\displaystyle \frac{1}{\sqrt{|N(v_j)|}} \sum_{v_m \in N(v_j)} \ket{v_m}$ & yes/no & \\
        \cline{1-4}
        \begin{tikzpicture}[baseline=0,MyStyle/.style={draw, minimum width=1.5em, minimum height=1.5em, outer sep=0pt}]
            \matrix (A) [matrix of math nodes, nodes={MyStyle, anchor=center}, column sep=-\pgflinewidth]
        {e_1 & e_2 & \cdots & \cdots \\};
            \addvmargin{1mm}
        \end{tikzpicture} & $\ldots$ & $\ldots$ & $\ldots$ &\\
        \hline
        \hline
        \begin{tikzpicture}[baseline=0,MyStyle/.style={draw, minimum width=1.5em, minimum height=1.5em, outer sep=0pt}]
            \matrix (A) [matrix of math nodes, nodes={MyStyle, anchor=center}, column sep=-\pgflinewidth]
        {e_1 & e_2 & \cdots & \cdots \\};
            \addvmargin{1mm}
        \end{tikzpicture} & $\ldots$ & $\ldots$ & $\ldots$ & \Block{3-1}{\textsc{Partial-Anchor Scheme}\\ \\ if $v_i \in T$: $\mathcal{O}(\sqrt{n_e})$ entries\\one for each $v_k \in T$\\ \\ if $v_i \not \in T$: no entries} & \Block{3-1}{\textsc{Full-Anchor Scheme}\\ \\ \\$\mathcal{O}(\sqrt{n_e})$ entries\\one for each $v_k \in t(v_i)$ \\ \\} \\
        \cline{1-4}
        \begin{tikzpicture}[baseline=0,MyStyle/.style={draw, minimum width=1.5em, minimum height=1.5em, outer sep=0pt}]
            \matrix (A) [matrix of math nodes, nodes={MyStyle, anchor=center}, column sep=-\pgflinewidth]
        {e_1 & e_2 & \cdots & \cdots \\};
            \addvmargin{1mm}
        \end{tikzpicture}
            & $\ket{v_k}$ & $\displaystyle  \frac{1}{\sqrt{|N(v_k)|}} \sum_{v_m \in N(v_k) } \ket{v_m}$ & yes & \\
        \cline{1-4}
        \begin{tikzpicture}[baseline=0,MyStyle/.style={draw, minimum width=1.5em, minimum height=1.5em, outer sep=0pt}]
            \matrix (A) [matrix of math nodes, nodes={MyStyle, anchor=center}, column sep=-\pgflinewidth]
        {e_1 & e_2 & \cdots & \cdots \\};
            \addvmargin{1mm}
        \end{tikzpicture} & $\ldots$ & $\ldots$ & $\ldots$ & \\
        \hline
        \hline
        \CodeAfter
            \SubMatrix{.}{3-3}{5-4}{\}}
            \SubMatrix{.}{6-3}{8-4}{\}}
    \end{NiceTabular}
    \caption{Schematic view of the entangling table of ESP $v_i$. The former portion is devoted to track entanglement within e-neighborhood (low-cost artificial links) and it is common in both \textit{Partial-Anchor} and \textit{Full-Anchor} schemes. The latter portion is devoted to maintain high-cost artificial links, and it differs depending on the adopted scheme.}
    \label{fig:06}
    \hrulefill
\end{figure*}

As a consequence, an arbitrary ESP $\ket{v_i}$ maintains an \textit{entangling routing table} with $\tilde{\mathcal{O}}(\sqrt{n_e})$ entries, as enforced by our \textit{compact} design principle. This table is organized as shown in Fig.~\ref{fig:06} and detailed below:
\begin{itemize}
    \item  $\mathcal{O}(\sqrt{n_e} \log{n_e})$ ``entries'' are dedicated to e-neighbors in $N(v_i)$, with each entry storing a certain number\footnote{The number of e-bits per entry is a design parameter, that depends on the number of communication qubits available at each ESP. Thus this choice is dictated by the hardware-specific complexity, characterizing the ESPs.} of e-bits shared with $v_j$ in $ N(v_i)$, along with:
    \begin{enumerate}
        \item[i)] the quantum address $\ket{v_j}$,
        \item[ii)] and the superposition of the quantum addresses $\{\ket{v_m}\}$ of the e-neighbors in $N(v_j)$.
    \end{enumerate}
    \item  If $v_i \in T$ then there exist $\mathcal{O}(\sqrt{n_e})$ additional ``entries'', storing a certain number of e-bits shared with each anchor ESP $v_k \in T$, together with:
    \begin{enumerate}
        \item[i)] the quantum address $\ket{v_k}$,
        \item[ii)] and the superposition of the quantum addresses of the e-neighbors in $N(v_k)$.
    \end{enumerate}
\end{itemize}

It is important to note that an artificial link between $v_i$ and $v_j$, representing a shared entangled state, requires that (at least) one ebit is stored at each node. Thus, whenever an ESP $v_i$ establishes a link with its e-neighbor $v_j$, there is an entry in $v_i$ quantum routing table as well as a corresponding entry in $v_j$ quantum routing table.
However, the neighbor sets of $v_i$ and $v_j$ may differ: the entangling cost function $w(\cdot,\cdot)$ may induce the selection of $v_i$ among the top $\sqrt{n_e}$ entanglement-neighbors of $v_j$, i.e., $v_i \in N(v_j)$, but the reverse is not true, i.e., $v_j \notin N(v_i)$. This asymmetry may arise, since the neighbors for $v_i$ and $v_j$ could be constructed independently and possibly from partial  views of the network.
To preserve routing consistency and full usability of the links established by $v_j$, node $v_i$ must be aware of such asymmetry. This motivates the introduction of the \textit{reverse neighborhood} of node $v_i$, defined as $R(v_i) \eqdef \bigcup \big\{ v_j \in V : v_i \in N(v_j) \big\}$. In practice this means that, while the primary entries in $v_i$ entangling table are determined by its own neighbor-set $N(v_i)$ and utilized for the packet forwarding at $v_i$, the table must also accommodate auxiliary entries for $R(v_i)$, to enable correct handling of routing requests terminating at or passing through $v_i$ due to its inclusion in another node’s neighbor-set.
Thus, the number of entries at $v_i$ scales as $|R(v_i) \cup N(v_i)|$. 
There exist some degenerate cases, where the cardinality of $|R(v_i)|$ is not upper-bounded by $\tilde{\mathcal{O}}(\sqrt{n_e})$, such as when the entangling cost $w(\cdot,\cdot)$ does not satisfy the axiomatic properties of a metric given in \eqref{eq:11}-\eqref{eq:14} or when the cost induces strong centralization, with a few privileged nodes being entangled-efficient for a disproportionately large number of others. In the following, we reasonably assume that such degenerate cases can be properly handled by fairness policies enforced by the EDC. These policies can induce a maximum number of entangling table entries per node, prioritizing fairness and scalability. 

\vspace{3pt}

\textbf{Quantum Routing Logic:}\\
We are ready now to present the logic underlying our compact quantum routing protocol for the partial-anchor scheme, achieving entanglement-stretch $\mathscr{ES}$ upper bounded by 5, as proved in Lemma~\ref{lem:01}.
More in detail, end-to-end entanglement between an ``initiating'' ESP, say node $v_i$, and a ``target'' ESP $v_d$ proceeds through one of the following three cases, examined in order.
\begin{itemize}
    \item{\textit{Case I}.} $v_d$ belongs to $N(v_i)$, or equivalently if $v_i \in T$, $v_d$ may belong to $T$ as well. This case can be checked by simply searching for quantum address $\ket{d}$ within the \textit{e-hop} field of the entangling table entries, by exploiting the orthogonality of the addresses. If this condition holds, then $v_i$ has already established entanglement with $v_d$, through the optimal entangling metric by design. Thus, an artificial link exists with unitary entangling stretch, i.e., $\mathscr{ES}(i,d)=1$.
    \item{\textit{Case II}.} This case is illustrated in Fig.~\ref{fig:07a}. If \textit{Case I} does not hold, the targeted ESP $v_d$ may still belong to the e-neighborhood of some $v_j \in N(v_i)$. This can be checked by searching for the quantum address $\ket{v_d}$ within the superposed quantum addresses stored in the \textit{e-neighborhood} field of the entangling table entries, by exploiting the \textit{quantum addressing splitting} functionality designed in Sec.~\ref{sec:4}. If $\ket{v_d}$ is found, by denoting with $v_j \in N(v_i)$ the \textit{e-hop} of $\ket{v_d}$ as in Fig.~\ref{fig:07a}, then both $ v_i , v_j $ and $ v_j , v_d $ are entangled through the optimal entangling metric by design. Thus, two artificial links with unitary stretch exist -- $\mathscr{ES}(i,j) = \mathscr{ES}(j,d) = 1$ -- and they enable end-to-end entanglement between $v_i$ and $v_d$ via entanglement swapping at $v_j$.
    \item{\textit{Case III}.} This case is represented in Fig.~\ref{fig:07b}. Whenever the previous two cases do not apply then, with a probability that can be made as close to 1 as wished accordingly to Property~\ref{proper:01}, there exists an anchor ESP, say $v_l \in T$, within the e-neighborhood of $v_i$\footnote{The underlying and non-restrictive hypothesis is that each anchor ESP reveals its anchor-status to its e-neighbors. In other words, if $v_l \in T$ belongs to the e-neighborhood $N(v_i)$, then $v_i$ is aware that $v_l$ is an anchor. This information is required to allow forwarding decisions that leverage the entanglement already shared among the anchors. Importantly, this does not entail any global coordination, only local exchanges during the entanglement establishment phase, achieved by embedding such metadata into the exchanged quantum packets, carrying also the superposed state encoding the identities of the neighbors of each e-hop.}. By design, $v_l$ has already established entanglement with any other anchor in $T$, including some anchor $v_k \in T$ that belongs to the e-neighborhood of $v_d$. 
    This can be verified by searching for the quantum address $\ket{v_d}$ in the \textit{e-neighborhood} fields associated with the e-hops field of $v_l \in T$. As a consequence, there exist three artificial links characterized by unitary entangling stretch -- i.e. $\mathscr{ES}(i,l) = \mathscr{ES}(l,k) = \mathscr{ES}(k,d) = 1$ -- and thus end-to-end entanglement between $v_i$ and $v_d$ can be straightforwardly obtained by entanglement swapping at both $v_l$ and $v_k$\footnote{\label{foot:03} There exists some special configurations of \textit{Case III} -- such as when $v_i \in T$ as shown in Fig.~\ref{fig:07c} or when $v_d \in T$ -- that yield a tighter upper bound for the entangling stretch, i.e., $\mathscr{ES} < 3$. In particular, when both $v_i,v_d \in T$, the configuration actually falls under Case I, since anchor ESPs are directly entangled by design, resulting in unitary stretch $\mathscr{ES} = 1$.}.
\end{itemize}

\begin{remark}
    \label{rem:06}
    The configuration described in \textit{Case III} substantiates the key insight anticipated at the beginning of Sec.~\ref{sec:3.3}. Specifically, with high probability, \textit{every ESP is at most 3-hop-entanglement distant from any other ESP} within the overlay topology, and this confirms the strategic role of the anchors as entanglement hubs. Importantly, this 3-hop value represents a worst-case scenario, derived under the assumption that both $v_i$ and $v_d$ are neither anchor nor $k$-closest neighbor nodes, thus without direct entanglement. In many practical configurations, the hop-entanglement can be strictly smaller. Nevertheless, this worst-case guarantee offers strong evidence of the routing performance, achieved with only sublinear memory overhead and local decision-making. Thus, it reinforces the effectiveness of the anchor-based overlay design.
\end{remark}

\begin{remark}
    \label{rem:07}
    Although we fairly impose an \emph{optimal entangling metric} $w(\cdot,\cdot)$ for the selection of the e-neighbors, by ensuring that each ESP maintains entanglement with its most favorable peers according to a selected performance criterion, the entangling stretch $\mathscr{ES}$ of the proposed routing scheme is not necessarily equal to one. This is the consequence of two structural relaxations deliberately introduced in our design to preserve the \textit{compactness and scalability} of our protocol.\\
    First, we \textit{relax the requirement of a fully connected entangled mesh among the ESPs}. While such a mesh would guarantee unit stretch for all the aforementioned cases, even in \textit{Case III}, maintaining all-to-all entanglement among ESPs would be practically unsustainable, due to the complexity related to the hardware and control overhead. Instead, our protocol relies on a sparse backbone, built through localized decisions and bounded connectivity.\\
    Second, but most importantly, we do \textit{not optimize the construction of the anchor set $T$}, as mentioned above. The anchor ESPs are selected through the \textit{randomized, distributed algorithm}, described in Remark~\ref{rem:05}. While this approach requires no global knowledge, it may yield a suboptimal anchor placement from the perspective of minimizing end-to-end entangling stretch. In this light and in agreement with Remark~\ref{rem:05}, it is reasonable to assume the nodes in $T$ equipped with \textit{enhanced quantum hardware capabilities}. Under this assumption, it may be more appropriate to construct the anchor set \textit{deterministically}. Such a deterministic approach, potentially combined with a joint optimization of the entangling cost metric and the anchor topology, could further improve the routing efficiency, by achieving even lower stretch, even in scenarios corresponding to  \textit{Case III}. However, a detailed investigation of this joint optimization strategy falls beyond the scope of this paper. We leave it as future work.
\end{remark}

The entangling stretch of the proposed partial-anchor scheme is upper-bounded formally in the following Lemma.

\begin{lem}
    \label{lem:01}
    By denoting with $v_i$ and $v_d$ the ESPs exhibiting the worst-case entangling stretch defined in \eqref{eq:19}, the stretch of the corresponding entangling path is upper bounded as follow:
    \begin{align}
        \label{eq:28}
        \mathscr{ES} &\leq c = \frac{w\big( (i,d) \oplus (i,d) \oplus (i,d) \oplus (i,d) \oplus (i,d) \big)}{w(i,d)} \nonumber \\
            & \eqdef \frac{w\left( \bigoplus_{5} (i,d) \right)}{w(i,d)}.
    \end{align}
    For additive metrics, i.e., when the composition operator is such that $w(a \oplus b) = w(a) + w(b)$,  \eqref{eq:28} simplifies to:
    \begin{equation}
        \label{eq:29}
        \mathscr{ES} \leq c = 5. 
    \end{equation}    
    Conversely, for concave metrics where the composition operator is defined as $\oplus \eqdef \min$, the stretch in  \eqref{eq:28} becomes unitary, i.e., $\mathscr{ES} = 1$.
    \begin{IEEEproof}
        See Appendix~\ref{app:02}.
    \end{IEEEproof}
\end{lem}

\subsection{Full-Anchor Scheme}
\label{sec:3.4}

Here we design a second quantum routing scheme, which relies on enhanced ESP capabilities to establish long-range artificial links. To this aim, while the definition of e-neighborhood in Def.~\ref{def:05} remains unchanged, we introduce the following design principle, as an alternative to Design Principle~\ref{Des:05}.

\begin{subdesign}[\textbf{Full-Anchor Scheme}]
    \label{Des:06}
    The design of the quantum routing protocol requires each ESP to proactively generate and sustain both the two ``species'' of artificial links:  short-range (i.e., low-cost) and long-range (i.e., high-cost) links.
\end{subdesign} 

Accordingly, in this second routing scheme, each ESP serves as anchor. To emphasize this aspect, we termed it \textit{full-anchor scheme}. However, to avoid unsustainable routing-table growth, we induce a sparse tracking paradigm, where each ESP monitors only a subset, randomly selected, of other ESPs. A naive design of this scheme may result in pathological scenarios, such as for instance, all ESPs tracking the same subset, leaving parts of the overlay unreachable. Consequently, to ensure that the sparse tracking preserves full-network coverage, we leverage the result for extended dominating sets, introduced in the previous subsection. To formalize this approach, the following definitions are needed.

\begin{defin}[\textbf{Tracked-Sets}]
    \label{def:07}
    A collection of disjoint subsets $\mathcal{T} = \left\{ T_1, \ldots, T_{\sqrt{n_e}} \right\}$ is said to form a partition of the ESP set $V_2$ if:
    \begin{equation}
        \label{eq:30}
         \quad \bigcup_{i} T_i = V_2 \; \wedge \; T_i \cap T_j = \emptyset.
    \end{equation}
In the following, we refer to the elements in $\mathcal{T}$ as \textit{tracked-sets}. We assume that the mapping function $ V \longrightarrow \mathcal{T}$, determining the partition of the ESP set $V_2$, is globally known by the nodes.
\end{defin}

In the following, for the sake of simplicity we assume the tracked-sets are constructed via an arbitrary flat partitioning of the ESP set $V_2$. One natural choice is to partition nodes based on their quantum addresses, i.e., a node with quantum address $\ket{i}$ is assigned to the $j$-th tracking-set $T_j$ if and only if $(j-1) \sqrt{n_e} < i \leq j \sqrt{n_e}$.

\begin{remark}
    The aforementioned flat partitioning has the desirable property of being independent of the network topology and yet it guarantees that each tracked-set $T_j$ contains at most $\sqrt{n_e}$ nodes, i.e. $T_j \leq \sqrt{n_e}$ for any $j$. Nevertheless, any alternative flat partitioning that respects this cardinality constraint is equally valid.
\end{remark}

We now define the \textit{tracking-nodes}.

\begin{defin}[\textbf{Tracking-nodes}]
    \label{def:08}
    Each ESP $v_i \in V_2$ tracks only one tracked-set $T_j \in \mathcal{T}$, chosen uniformly at random. This set is denoted as $t(v_i)$:
    \begin{equation}
        \label{eq:31}
        \forall \, v_i \in V_2 : \exists! \, T_j \in \mathcal{T} \text{ s.t. } t(v_i) = T_j
    \end{equation}
    Any node $v_d \in t(v_i)$ is referred to as \textit{one of the nodes tracked by $v_i$}.
\end{defin}

The key observation is that the union of tracked-sets originating from the e-neighborhood of any given ESP must collectively cover the entire ESP set. This ensures reachability and global coverage under sparse tracking regime.
To enforce this coverage, we leverage again the result on extended dominating sets \cite{Lov-75}, introduced in the previous section. Indeed, by enforcing the same parametrization of the e-neighborhood cardinality $k = (1+m) \sqrt{n_e} \log n_e$, we ensure the following property with high-probability:

\begin{proper}
    \label{proper:02}
    Any ESP $v_i$ has in its neighborhood at least one node tracking any other ESP w.h.p., i.e.:
    \begin{equation}
        \label{eq:32}
        P \big( \not\exists \, v_j \in N(v_i) : v_d \in t(v_j) \big)= \mathcal{O}(n_e^{1-m}), \,\, \forall \, v_i, v_d \in V_2.
    \end{equation}
\end{proper}

This property guarantees that, although each ESP tracks only a limited number of other ESPs, the collective coverage provided by the e-neighborhood is sufficient to ensure full overlay reachability.

\begin{remark}
    \label{rem:08}
    As noted in Remark~\ref{rem:05}, a similar observation applies here as well: the specific construction of the tracked sets lies outside the scope of this work. Our primary objective is to demonstrate that Property~\ref{proper:02} can be ensured with high probability, through appropriate parametrization. Although more optimized constructions are possible, our analysis deliberately avoids assuming any particular structure in the tracked set formation. As such, the presented results should be interpreted as worst-case guarantees.
\end{remark}

With the above in mind, the full-anchor scheme relies on the following:
\begin{quote}
    \textit{each ESP $v_i$ proactively establishes and maintains entanglement with any node within its tracked-set $t(v_i)$,}
\end{quote}
where entanglement is established using the optimal end-to-end entangling metric in Def.~\ref{def:03}. Accordingly, $v_i$ and any $v_j \in t(v_i)$ are connected by an artificial link within the artificial topology characterized by unitary entangling stretch, i.e., $\mathscr{ES}(i,j) = 1$. Furthermore:
\begin{quote}
    \textit{each ESP $v_i$ proactively establishes and maintains entanglement with any ESP within its e-neighborhood $N(v_i)$,}
\end{quote}
again via the optimal end-to-end entangling metric in Def.~\ref{def:03}. Consequently, each artificial link between $v_i$ and any $v_k \in N(v_i)$ also exhibits a unitary entangling stretch, i.e., $\mathscr{ES}(i,k) = 1$.

As a consequence, an arbitrary ESP $\ket{v_i}$ maintains an \textit{entangling routing table} with $\tilde{\mathcal{O}}(\sqrt{n_e})$ entries, as enforced by our \textit{compact} design principle. This table is organized as shown in Fig.~\ref{fig:06}. The distinguishing factor between the \textit{partial-anchor scheme} and the \textit{full-anchor scheme} is the latter part of the quantum routing table devoted to high-cost artificial links. Specifically, the \textit{full-anchor scheme} allows each ESP, rather than only the nodes in $T$, to proactively establish and maintain such links. This maps into an ``enhanced'' connectivity within the artificial topology, which enables our compact quantum routing protocol to guarantee entangling stretch $\mathscr{ES}$ upper bounded by 3, as in \eqref{eq:21}, with at most $\tilde{\mathcal{O}}(\sqrt{n_e})$ qubits to be stored at each node.

\begin{figure*}
	\centering
	\begin{minipage}[c]{0.45\textwidth}
		\centering
        \begin{adjustbox}{width=\columnwidth}
            \begin{tikzpicture}[->]
                \node [circle, draw, fill=myBlue, text=white] (i) at (0,0) {$v_{i}$};
                \draw[fill=none,color=myBlue](0,0) circle (2.9) node [myBlue,yshift=-2.5cm] {$N(v_i)$};
                
                \node [circle, draw, fill=myNeighborSet, text=white] (j) at (2,-1.5) {$v_{j}$};
                \node[myNeighborSet] at (3,-2) {$v_j \in N(v_i)$};

                \node [circle, draw, fill=myTrackingSet, text=white] (k) at (4,2) {$v_{k}$};
                \node[myTrackingSet] at (3.2,2.6) {$v_k \in T : v_d \in N(v_k)$};

                \node [circle, draw, fill=myTrackingSet, text=white] (l) at (-1,1.5) {$v_{l}$};

                \node [circle, draw, fill=myBlue, text=white] (d) at (4,0) {$v_{d}$};
                \node[myBlue] at (5,-0.5) {$v_d \in N(v_j)$};
                
                \draw[very thick,loosely dashed,<->,myPurple] (i) edge node[xshift=-25pt,yshift=-5pt] {\textsc{\small $\mathscr{ES}(i,j) = 1$}} (j);
                
                \draw[very thick,loosely dashed,<->,myPurple] (j) edge node[xshift=25pt,yshift=-5pt] {\textsc{\small $\mathscr{ES}(j,d) = 1$}} (d);

                \draw[very thick,loosely dashed,bend left,myPurple] (i) edge node[above,yshift=2pt] {\textsc{\small $\mathscr{ES}(i,d) \leq 3$}} (d);
            \end{tikzpicture}
        \end{adjustbox}
    	\subcaption{Case II: there exists $v_j \in N(v_i)$ so that $v_d \in N(v_j)$.}
		\label{fig:07a}
	\end{minipage}
	\hspace{0.04\textwidth}
	\begin{minipage}[c]{0.45\textwidth}
		\centering
        \begin{adjustbox}{width=\columnwidth}
            \begin{tikzpicture}[->]
                \node [circle, draw, fill=myBlue, text=white] (i) at (0,0) {$v_{i}$};
                \draw[fill=none,color=myBlue](0,0) circle (2.9) node [myBlue,yshift=-2.5cm] {$N(v_i)$};
                
                \node [circle, draw, fill=myNeighborSet, text=white] (j) at (2,-1.5) {$v_{j}$};

                \node [circle, draw, fill=myTrackingSet, text=white] (k) at (4,2) {$v_{k}$};
                \node[myTrackingSet] at (3.2,2.6) {$v_k \in T : v_d \in N(v_k)$};

                \node [circle, draw, fill=myTrackingSet, text=white] (l) at (-1,1.5) {$v_{l}$};
                \node[myTrackingSet] at (0,2.1) {$v_l \in N(v_{i}) : v_l \in T$};

                \node [circle, draw, fill=myBlue, text=white] (d) at (4,0) {$v_{d}$};
                \node[myBlue] at (5,-0.5) {$v_d \in N(v_k)$};
                
                \draw[very thick,loosely dashed,<->,myPurple] (i) edge node[xshift=-25pt,yshift=-5pt] {\textsc{\small $\mathscr{ES}(i,l) = 1$}} (l);

                \draw[very thick,loosely dashed,<->,myPurple] (l) edge node[xshift=10pt,yshift=-10pt] {\textsc{\small $\mathscr{ES}(l,k) = 1$}} (k);

                \draw[very thick,loosely dashed,<->,myPurple] (k) edge node[xshift=30pt,yshift=0pt] {\textsc{\small $\mathscr{ES}(k,d) = 1$}} (d);

                \draw[very thick,loosely dashed,bend right,myPurple] (i) edge node[above,yshift=2pt] {\textsc{\small $\mathscr{ES}(i,d) \leq 5$}} (d);
            \end{tikzpicture}
        \end{adjustbox}
    	\subcaption{Case III: there exists $v_l,v_k \in T$ so that $v_l \in N(v_i)$ and $v_k \in N(v_d)$.}
		\label{fig:07b}
	\end{minipage}
	\vspace{0.07\textwidth}    
	\begin{minipage}[c]{0.45\textwidth}
		\centering
        \begin{adjustbox}{width=\columnwidth}
            \begin{tikzpicture}[->]
                \node [circle, draw, fill=myTrackingSet, text=white] (i) at (0,0) {$v_{i}$};
                \draw[fill=none,color=myTrackingSet](0,0) circle (2.9) node [myBlue,yshift=-2.5cm] {$N(v_i)$};
                
                \node [circle, draw, fill=myNeighborSet, text=white] (j) at (2,-1.5) {$v_{j}$};
    
                \node [circle, draw, fill=myTrackingSet, text=white] (k) at (4,2) {$v_{k}$};
                \node[myTrackingSet] at (3.2,2.6) {$v_k \in T : v_d \in N(v_k)$};

                \node [circle, draw, fill=myNeighborSet, text=white] (l) at (-1,1.5) {$v_{l}$};

                \node [circle, draw, fill=myBlue, text=white] (d) at (4,0) {$v_{d}$};
                \node[myBlue] at (5,-0.5) {$v_d \in N(v_k)$};
                
                \draw[very thick,loosely dashed,<->,myPurple] (i) edge node[xshift=-30pt,yshift=5pt] {\textsc{\small $\mathscr{ES}(i,k) = 1$}} (k);

                \draw[very thick,loosely dashed,<->,myPurple] (k) edge node[xshift=25pt,yshift=-5pt] {\textsc{\small $\mathscr{ES}(k,d) = 1$}} (d);

                \draw[very thick,loosely dashed,bend right,myPurple] (i) edge node[above,yshift=2pt] {\textsc{\small $\mathscr{ES}(i,d) \leq 3$}} (d);
            \end{tikzpicture}
        \end{adjustbox}
    	\subcaption{Case IV as particular configuration of case III where either $v_i$ belongs to $T$.}
		\label{fig:07c}
	\end{minipage}
    \caption{Partial-Anchor Scheme: schematic view of the logic underlying our compact quantum routing protocol for establishing end-to-end entanglement between ``initiating'' ESP $v_i$ and ``target'' ESP $v_d$. Tracking nodes in $T$ are denoted in purple, artificial links are denoted with red arrows, and circles around quantum nodes denote their e-neighborhood.}
    \label{fig:07}
    \hrulefill
\end{figure*}
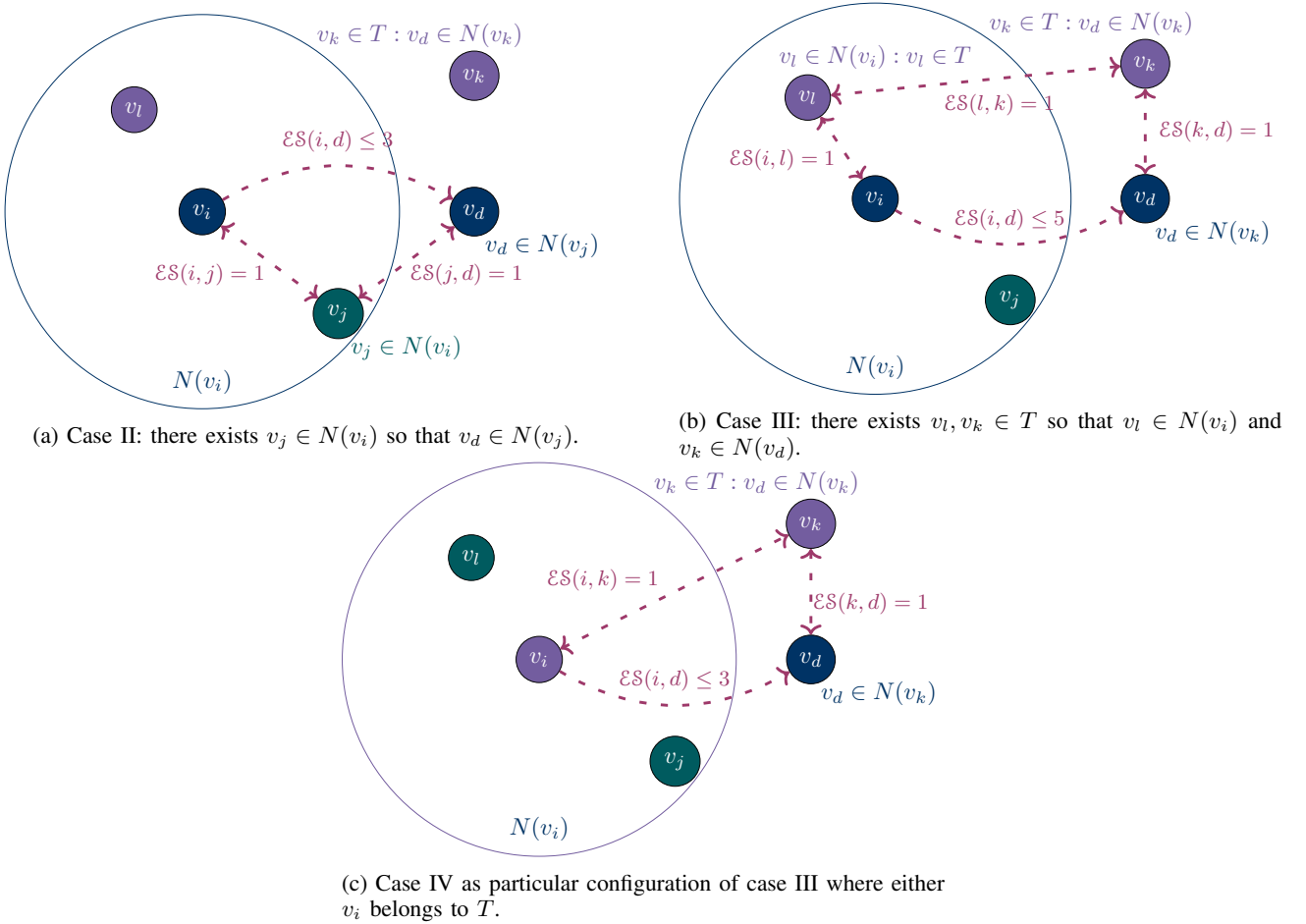

\vspace{3pt}
\textbf{Quantum Routing Logic:}\\
We are ready now to describe the logic underlying \textit{full-anchor scheme}. More in detail, end-to-end entanglement between ``initiating'' ESP $v_i$ and ``target'' ESP $v_d$ proceeds through one of the following two cases, examined in order.
\begin{itemize}
    \item \textit{Case I}. $v_d$ may belong to $N(v_i)$ or $t(v_i)$; if so, similarly to \textit{Case I} of the \textit{Partial-Anchor Scheme}, then $v_i$ already established some entanglement with $v_d$ through the optimal entangling metric by design. Thus, an artificial link exists with unitary entangling stretch: $\mathscr{ES}(i,d)=1$.
    \item \textit{Case II}. It is similar to the case represented in Fig.~\ref{fig:07a}. Whenever the first case does not hold, according to Property~\ref{proper:02} which holds w.h.p., $v_d$ belongs to the e-neighborhood of at least one node in $N(v_i)$. This can be checked by searching for the quantum address $\ket{v_d}$ within the superposed quantum addresses stored in the \textit{e-neighborhood} field of the entangling table entries, by exploiting the \textit{addressing splitting} functionality designed in Sec.~\ref{sec:4}. If $\ket{v_d}$ is found, by denoting with $v_j \in N(v_i)$ the \textit{e-hop} of $v_d$ as in Fig.~\ref{fig:07a}, then both $v_i,v_j$ and $ v_j , v_d $ are entangled with unitary stretch by design. Thus, two artificial links with unitary stretch exist. As a consequence, end-to-end entanglement between $v_i$ and $v_d$ can be straightforwardly obtained by entanglement swapping at $v_j$.
\end{itemize}

\begin{remark}
    \label{rem:09}
    Similarly to the observation in Remark~\ref{rem:06}, the configuration described in \textit{Case II} substantiates the key insight anticipated at the beginning of Sec.~\ref{sec:3.3}. Specifically, with high probability, \textit{every ESP is at most at 2 entanglement-hops from any other ESP} in the overlay topology.
\end{remark}

\begin{lem}
    \label{lem:02}
    By denoting with $v_i$ and $v_d$ the ESPs exhibiting the worst-case entangling stretch defined in \eqref{eq:19},  the stretch of the corresponding entangling path is upper bounded as follow:
    \begin{align}
        \label{eq:33}
        \mathscr{ES} &\leq c = \frac{w\big( (i,d) \oplus (i,d) \oplus (i,d) \big)}{w(i,d)} \eqdef \frac{w\left( \bigoplus_{3} (i,d) \right)}{w(i,d)}
    \end{align}
    which, for additive metrics \eqref{eq:28}, simplifies to:
    \begin{equation}
        \label{eq:34}
         \mathscr{ES} \leq c = 3.
    \end{equation}    
    Conversely, for concave metrics ($\oplus \eqdef \min$) the stretch in \eqref{eq:25} becomes unitary, i.e., $\mathscr{ES} = 1$.
    \begin{IEEEproof}
        The proof follows by adopting the same reasoning as in App.~\ref{app:02}
    \end{IEEEproof}
\end{lem}

\section{Quantum Address Splitting via Schr\"odinger's Oracle}
\label{sec:4}

\begin{figure*}
    \centering
    \begin{adjustbox}{width=1\linewidth}
        \begin{quantikz}
            \lstick[4]{\shortstack{\sc entry\\\sc label \\\sc register}} \slice{\shortstack{\sc standard \\ \sc Grover\\ \sc inizialization}} & \gate{H} & \qw & \ldots\ \slice{\shortstack{\sc Schr\"odinger's \\\sc Grover\\ \sc iteration}} & [1.5cm] \gategroup[22,steps=6,style={dashed,rounded corners,inner xsep=2pt},background,label style={label
position=below,anchor=north,yshift=-0.2cm}]{{\shortstack{\sc coherent-controlled oracle\\\sc for first entry}}} \qw & \ctrl[open]{13} & \qw & \qw & \ctrl[open]{13} & \qw & \ldots \ & \gategroup[22,steps=6,style={dashed,rounded corners,inner xsep=2pt},background,label style={label
position=below,anchor=north,yshift=-0.2cm}]{{\shortstack{\sc coherent-controlled oracle\\\sc for eleventh entry}}} \qw & \ctrl{13} & \qw & \qw & \ctrl{13} & \qw & \ldots \ & \gate[4]{\shortstack{\scshape{standard}\\\scshape{Grover}\\\scshape{Diffusion}}} \slice{\shortstack{\sc next \\ \sc Schr\"odinger's Grover\\ \sc iteration}} & \ldots\ & \meter[4]{} \\
            & \gate{H} & \qw & \ldots\ & \qw & \ctrl[open]{20} & \qw & \qw & \ctrl[open]{20} & \qw & \ldots \ & \qw & \ctrl[open]{20} & \qw & \qw & \ctrl[open]{20} & \qw & \ldots \ & \qw & \ldots\ & \qw \\
            & \gate{H} & \qw & \ldots\ & \qw & \ctrl[open]{19} \qw & \qw & & \ctrl[open]{19} & \qw & \ldots \ & \qw & \ctrl{19} & \qw & \qw & \ctrl{11} & \qw & \ldots \ & \qw & \ldots\ & \qw \\
            & \gate{H} & \qw & \ldots\ & \qw & \ctrl[open]{18} & \qw & \qw & \ctrl[open]{18} & \qw & \ldots \ & \qw & \ctrl{18} & \qw & \qw & \ctrl{10} & \qw & \ldots \ & \qw & \ldots\ & \qw \\
            \wave & & & & & & & & & & & & & & & & & & & & \\
            \lstick[3]{\shortstack{\sc first\\ \sc superposed\\ \sc address\\$\Ket{A_0^1}$}} & \qw & \qw & \ldots\ & \gate{X^{t_{1}}} & \ctrl[open]{16} & \gate{X^{t_{1}}} & \qw & \qw & \qw & \ldots\ & \qw & \qw & \qw & \qw & \qw & \qw & \ldots\ & \qw & \ldots\ & \qw \rstick[7]{\sc $f(n_e)$ superposed\\\sc addresses\\\sc representing\\\sc the $f(n_e)$ disjoint\\\sc partitions of the \\ \sc e-neighborhood\\\sc of the e-hop\\\sc associated with \\ \sc the first entry} \\
            & \qw & \qw & \ldots\ & \gate{X^{t_{2}}} & \ctrl[open]{15} & \gate{X^{t_{2}}} & \qw & \qw & \qw & \ldots\ & \qw & \qw & \qw  & \qw & \qw & \qw & \ldots\ & \qw & \ldots\ & \qw \\
            & \qw & \qw & \ldots\ & \gate{X^{t_{3}}} & \ctrl[open]{14} & \gate{X^{t_{3}}} & \qw & \qw & \qw & \ldots\ & & \qw & \qw  & \qw & \qw & \qw & \ldots\ & \qw & \ldots\ & \qw \\
            \wave & & & & & & & & & & & & & & & & & & & \\
            \lstick[3]{\shortstack{$f(n_e)$-th\\ \sc superposed\\ \sc address\\$\Ket{A_0^{f(n_e)}}$}} & \qw & \qw & \ldots & \qw & \qw & \qw & \gate{X^{t_{1}}} & \ctrl[open]{12} & \gate{X^{t_{1}}} & \ldots & \qw & \qw & \qw & \qw & \qw & \qw  & \ldots\ & \qw & \ldots\ & \qw \\
            & \qw & \qw & \ldots & \qw & \qw & \qw & \gate{X^{t_{2}}} & \ctrl[open]{11} & \gate{X^{t_{2}}} & \ldots & \qw & \qw & \qw & \qw & \qw & \qw  & \ldots\ & \qw & \ldots\ & \qw \\
            & \qw & \qw & \ldots & \qw & \qw & \qw & \gate{X^{t_{3}}} & \ctrl[open]{10} & \gate{X^{t_{3}}} & \ldots & \qw & \qw & \qw & \qw & \qw & \qw  & \ldots\ & \qw & \ldots\ & \qw \\
            \wave & & & & & & & & & & & & & & & & & & & & \\
            \lstick[3]{\shortstack{\sc first\\ \sc superposed\\ \sc address}\\$\Ket{A_{10}^1}$} & \qw & \qw & \ldots\ & \qw & \qw & \qw & \qw & \qw & \qw & \ldots\ & \gate{X^{t_{1}}} & \ctrl[open]{8} & \gate{X^{t_{1}}} & \qw & \qw & \qw & \ldots\ & \qw & \ldots\ & \qw \rstick[7]{\sc $f(n_e)$ superposed\\\sc addresses\\\sc representing\\\sc the $f(n_e)$ disjoint\\\sc partitions of the \\ \sc e-neighborhood\\\sc of the e-hop\\\sc associated with \\ \sc the eleventh entry} \\
            & \qw & \qw & \ldots\ & \qw & \qw & \qw & \qw & \qw & \qw & \ldots\ & \gate{X^{t_{2}}} & \ctrl[open]{7} & \gate{X^{t_{2}}} & \qw & \qw & \qw & \ldots\ & \qw & \ldots\ & \qw \\
            & \qw & \qw & \ldots\ & \qw & \qw & \qw & \qw & \qw & \qw & \ldots\ & \gate{X^{t_{3}}} & \ctrl[open]{6} & \gate{X^{t_{3}}} & \qw & \qw & \qw & \ldots\ & \qw & \ldots\ & \qw \\
            \wave & & & & & & & & & & & & & & & & & & & \\
            \lstick[3]{\shortstack{$f(n_e)$-th\\ \sc superposed\\ \sc address\\$\Ket{A_{10}^{f(n_e)}}$}} & \qw & \qw & \ldots & \qw & \qw & \qw & \qw & \qw & \qw & \ldots & \qw & \qw & \qw & \gate{X^{t_{1}}} & \ctrl[open]{4} & \gate{X^{t_{1}}} & \ldots\ & \qw & \ldots\ & \qw \\
            & \qw & \qw & \ldots & \qw & \qw & \qw & \qw & \qw & \qw & \ldots & \qw & \qw & \qw & \gate{X^{t_{2}}} & \ctrl[open]{3} & \gate{X^{t_{2}}} & \ldots\ & \qw & \ldots\ & \qw \\
            & \qw & \qw & \ldots & \qw & \qw & \qw & \qw & \qw & \qw & \ldots & \qw & \qw & \qw & \gate{X^{t_{3}}} & \ctrl[open]{2} & \gate{X^{t_{3}}} & \ldots\ & \qw & \ldots\ & \qw \\
            \wave & & & & & & & & & & & & & & & & & & & & \\            %
            \lstick{\shortstack{\sc Grover\\ \sc ancilla}} & \gate{X} & \gate{H} & \ldots\ & \qw & \targ{} & \qw & \qw & \targ{} & \qw & \ldots\ & \qw & \targ{} & \qw & \qw & \targ{} & \qw & \ldots\ & \qw & \ldots\ & \qw
        \end{quantikz}
    \end{adjustbox}
    \caption{Quantum circuit representing the oracle of the Grover's search algorithm for quantum address $\ket{v_d}$, with binary coefficient $\{ t_i \}$ given in \eqref{eq:40}. The labels of the $n_T= \sqrt{n_e} \log{n_e}$ entries of the quantum routing table are encoded within the entry register -- represented in the figure as four qubits rather than $\left \lceil \log_2{ \left( \sqrt{n_e} \log{n_e} \right)} \right \rceil$ qubits for the sake of simplicity. The oracle performs a phase flip on the quantum state representing the label of entry associated to e-hop $\ket{v_j}$, whenever one of the superposed addresses, representing the disjoint e-neighborhoods of $\ket{v_j}$, contains the target quantum address $\ket{v_d}$. In the figure, we assumed for the sake of simplicity a number of entries equal to $16$. Thus, the entry register is $4$ qubits and each superposed address is $3$-qubits long. Furthermore, the top wire is the leftmost significant qubit, thus register state $\ket{1011}$ labels the eleventh entry.}
    \label{fig:08}
    \hrulefill
\end{figure*}
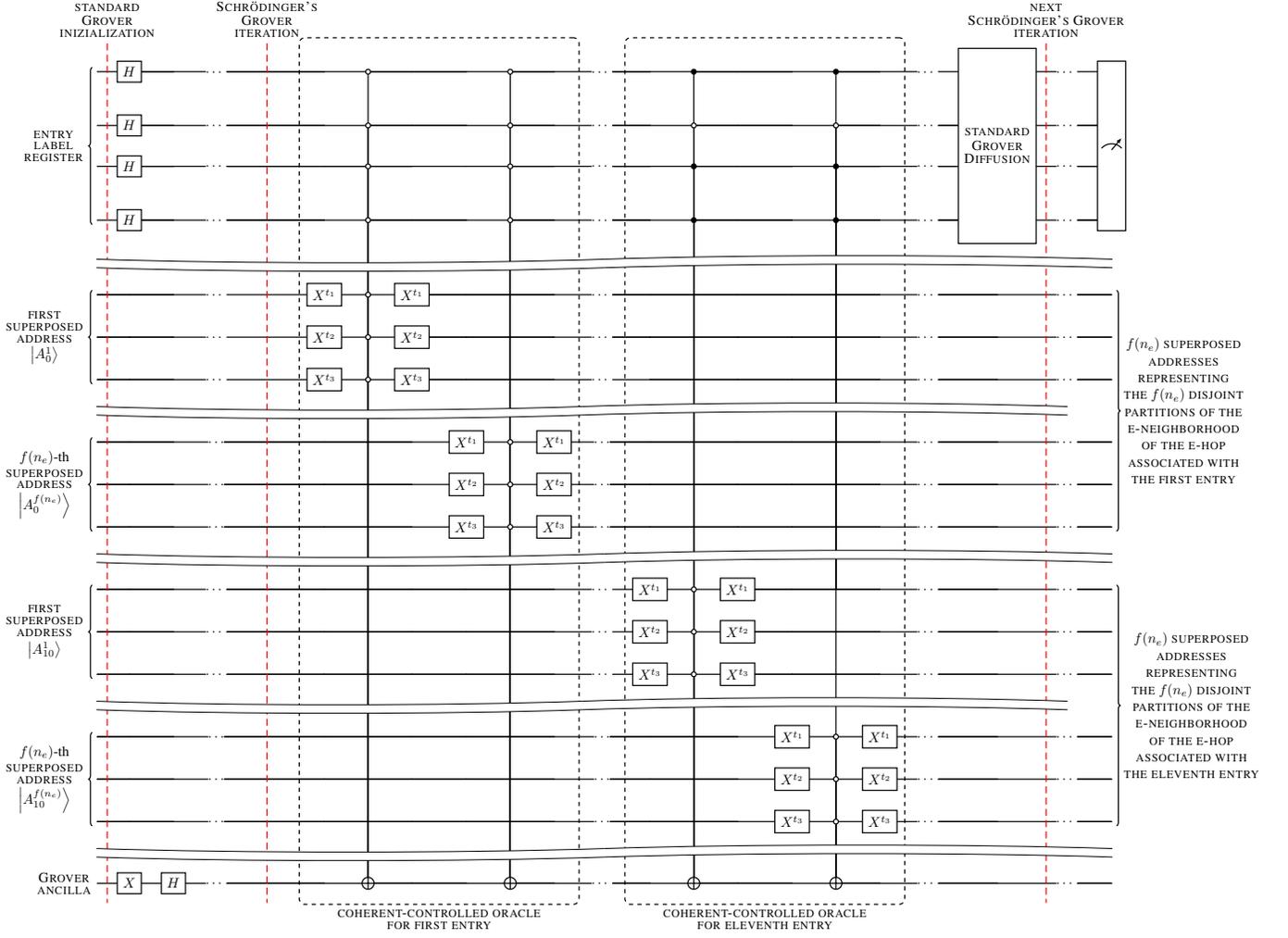

As detailed in the previous section, the proposed routing protocol, by exploiting superposed quantum addresses, requires the availability of a functionality we termed \textit{quantum address splitting}.

This functionality generalizes classical forwarding operations to quantum-superposed identifiers. Specifically, the quantum address splitting mechanism determines whether a given quantum address, say $\ket{v_d}$, is embedded within the superposed address stored in the \textit{e-neighborhood} field of a the quantum routing table entry, as illustrated in Fig.~\ref{fig:06}. And this check must be performed by satisfying two key constraints:
\begin{itemize}
    \item[i)] \textit{non-reliance on prior knowledge}: the addressing splitting functionality must not require any preliminary knowledge about the specific quantum addresses that have been superposed, since they depend on the topology via the e-neighborhood $N(\cdot)$;
    \item[ii)] \textit{non-destructive search}: the addressing splitting functionality should not alter any superposed address that does not contain the target address, $\ket{v_d}$, ensuring so that the superposed address remains available for subsequent queries involving different target addresses.
\end{itemize}

With the above two key constraints in mind, we propose a modified version of Grover’s quantum search algorithm \cite{Gro-96}, \textit{tailored for quantum-superposed data structures}. In our design, the oracle operates coherently on quantum-superposed entries and is thus referred to as the \textit{Schr\"odinger's oracle}. 
Specifically, the oracle is implemented via controlled quantum gates, where the control is provided by the quantum registers encoding the superposed addresses stored in the \textit{e-neighborhood} field of the routing table. The control condition is determined by the target quantum address $\ket{v_d}$ we aim to identify. 
As illustrated in Fig.~\ref{fig:08}, and detailed in the following, the oracle evaluates whether $\ket{v_d}$ is included in any superposed e-neighborhood associated with each routing entry. Specifically, it performs a phase inversion on the state representing the label of a given entry only when the corresponding superposed e-neighborhood contains the address $\ket{v_d}$.\\
This modified oracle causes the algorithm to evolve into a superposition of parallel search paths: one where the marked entry undergoes a phase flip, and another where it does not. Thus, the presence of the phase inversion becomes a coherent, quantum-controllable degree of freedom. This mechanism enables the selective identification of $\ket{d}$ without collapsing the superposition of unrelated addresses.

\subsection{Quantum Search Register}
\label{sec:4.1}

To fulfill the second key constraint, namely, the ability to perform non-destructive queries, we encode the labels of the routing table entries into a dedicated quantum register, which then undergoes the Grover search iterations. 
Formally, our search space consists of the\footnote{We introduce a more compact notation for the number of routing entries, with respect to the one adopted in Sec.~\ref{sec:3}, for the sake of brevity.} $n_{\text{T}}$ routing table entries, each uniquely identified by a label encoded as a computational basis state $\ket{y}$, with $y \in \{0,1\}^{N_{\text{T}}}$, and $N_{\text{T}} = \left\lceil \log_2(n_{\text{T}}) \right\rceil$. This quantum register, referred to as the \textit{entry label register}, holds the label of each entry. The register is initialized following the standard Grover preparation step, as shown in Fig.~\ref{fig:08}, resulting in an equal superposition of all the $n_{\text{T}}$ basis states:
\begin{equation}
    \label{eq:35}    
    \ket{\psi_0} = \frac{1}{\sqrt{n_{\text{T}}}} \sum_{y=0}^{n_{\text{T}} - 1} \ket{y}.
\end{equation}

\subsection{Quantum Superposed Address Registers}
\label{sec:4.2}
Let us consider the arbitrary entry labeled as $\ket{y}$ and denote the quantum address of its associated e-hop as $\ket{v_j}$.

The e-neighborhood $N(v_j)$ of node $v_j$ is partitioned by the node itself into $f(n_e)$ disjoint sub-e-neighborhoods, such that any given quantum node $v_k \in N(v_j)$ belongs to exactly one of these partitions. Each of these partitions is encoded as a quantum-superposed address, denoted as $\ket{A_j^l}$, and announced by $\ket{v_j}$. The superposition representing $l$-th partition of $N(v_j)$ is defined as follows:
\begin{equation}
    \label{eq:36}
    \ket{A_j^l} = \sqrt{\frac{f(n_e)}{\sqrt{n_e} \log n_e}} \sum_{v \in S_j^l} \ket{v},
\end{equation}
where $S_j^l$ is the set of nodes in the $l$-th partition of $N(v_j)$. Accordingly, each node $\ket{v}$ in $S_j^l$ appears in the quantum superposition $\ket{A_j^l}$ with amplitude $\sqrt{\frac{f(n_e)}{\sqrt{n_e} \log n_e}}$.

\begin{remark}
\label{rem:10}
    We highlight that $f(n_e)$ serves as a key tunable design parameter of our quantum address splitting functionality. Indeed, $f(n_e)$ allows us to drive the success probability of the Grover search via our Schr\"odinger's oracle toward one: the larger is $f(n_e)$, the higher is the success probability. 
\end{remark}

The $f(n_e)$ quantum addresses $\{\ket{A_j^l} \}$, jointly describing the e-neighborhood $N(v_j)$ of e-hop $\ket{v_j}$, control
our Schr\"odinger's oracle for the $j$-th entry label, as shown in Fig.~\ref{fig:08}.

\subsection{Schr\"odinger's Oracle}
\label{sec:4.3}
The Schr\"odinger's oracle $O_S$ for the e-hop $\ket{v_j}$ acts on the composite quantum system $\ket{\psi_0} \otimes \ket{A_j}$, with $\ket{A_j} \eqdef \ket{A_j^1} \otimes \ldots \otimes \ket{A_j^{f(n_e)}}$, as shown in equation \eqref{eq:38} shown\footnote{Where we omitted the superposed addressed for the e-neighborhoods of the other entries for the sake of notation simplicity and brevity. For the same reasons, we assume, without loss of generality, that $v_d$ is in $A^1_j$.} within next page. The amplitude associated with the target address in the first sub-e-neighborhood is denoted as $\sqrt{\alpha} \eqdef \braket{v_d | A^1_j} = \sqrt{\frac{f(n_e)}{\sqrt{n_e} \log n_e}}$.

\begin{figure*}
    \begin{align}
        \label{eq:38}
        &  O_S \left( \frac{1}{\sqrt{n_{\text{T}}}} \sum_{y=0}^{n_{\text{T}} - 1} \ket{y} \otimes \ket{A_j} \right) = O_S \left( \frac{1}{\sqrt{n_{\text{T}}}} \sum_{y=0}^{n_{\text{T}} - 1} \ket{y} \otimes \ket{A_j^1} \otimes \ldots \otimes \ket{A_j^l} \otimes \ldots \otimes \ket{A_j^{f(n_e)}} \right) = \nonumber \\
        & \quad \quad = \begin{cases}
            = \displaystyle \frac{1}{\sqrt{n_{\text{T}}}} \sum_{y=0}^{n_{\text{T}} - 1} \ket{y} \otimes \ket{A_j} &
                \text{if } v_d \not \in N(v_i)\\
            = \left\{
                \begin{aligned}
                    & = O_S \left( \frac{1}{\sqrt{n_{\text{T}}}} \sum_{y=0}^{n_{\text{T}} - 1} \ket{y} \otimes \left( \sqrt{1 - \alpha} \sum_{v \in S_j^1 \setminus \{v_d\}} \ket{v} + \sqrt{\alpha} \ket{v_d} \right) \otimes \ket{A_j^2} \otimes \ldots \otimes  \ket{A_j^{f(n_e)}} \right)=\\
                    & = \frac{1}{\sqrt{n_{\text{T}}}} \sum_{y=0,y \neq x}^{n_{\text{T}} - 1} \ket{y} \otimes \ket{A_j} +
                        \frac{1}{\sqrt{n_{\text{T}}}} \left( \sqrt{1 - \alpha} \ket{x} \otimes \sum_{v \in S_j^1, v \neq v_d} \ket{v} - \sqrt{\alpha} \ket{x} \otimes \ket{v_d} \right) \otimes \\
                    & \quad \quad \quad \quad \quad \quad \otimes \ket{A_j^2} \otimes \ldots \otimes  \otimes  \ket{A_j^{f(n_e)}}
                \end{aligned}
                \right. &
                \text{if } v_d \in A_j^1 \subset N(v_i)
        \end{cases}
    \end{align}
    \hrulefill
\end{figure*}
As evident from equation \eqref{eq:38} shown at the top of this page, whenever the target $\ket{v_d}$ does not belong to the e-neighborhood $N(v_j)$ of the e-hop $\ket{v_j}$, then our Schr\"odinger's oracle $O_S$ satisfies both the aforementioned constraints, by acting trivially on the entry label $\ket{x}$ (namely $\ket{x}$ is unchanged) and, crucially, leaving any non-matching superposed address unaltered. This ensures the non-destructive property of the search.
Conversely, whenever $\ket{v_d} \in N(v_j)$, then $O_S$ generates an inversion of the phase of the associated entry label $\ket{x}$. More precisely, $O_S$ entangles the entry label $\ket{x}$ with the ``hitting '' sub-e-neighborhood, inducing a conditional phase flip on $\ket{x}$. Thus, two computational evolutions are possible, one \textit{with} and the other \textit{without} the oracle phase inversion, as follows (by omitting all the irrelevant quantum subsystems):
\begin{align}
    \label{eq:39}
    & O_S \left( \frac{1}{\sqrt{n_{\text{T}}}} \ket{x} \otimes \left( \sqrt{1 - \alpha} \ket{v_d^{\perp}} + \sqrt{\alpha} \ket{v_d}  \right) \right) = \nonumber \\
    & \quad = \sqrt{\frac{1 - \alpha}{n_{\text{T}}}} \ket{x} \otimes \ket{v_d^{\perp}} - \sqrt{\frac{\alpha}{N_{\text{T}}}} \ket{x} \otimes \ket{v_d}.
\end{align}

\begin{remark}
    The Schr\"odinger's oracle $O_S$ allows Grover's algorithm to evolve in a coherent superposition of oracle-inverting and oracle-trivial computational dynamics. Further discussion in reported in Sec.~\ref{sec:5}.
\end{remark}

In Fig.~\ref{fig:08} we provide a straightforward, un-optimized circuit-level realization of the Schr\"odinger's oracle for $3$-qubits quantum addresses. The control conditions are determined by the binary coefficients $\{t_i\}$ driving the $X$-gates, derived from the classical representation of the target state $\ket{v_d}$:
\begin{equation}
    \label{eq:40}
    d = \sum_{i=0}^{N-1} b_i 2^i, \; b \in \{ 0 , 1 \},
\end{equation}
with $N$ defined in Def.~\ref{def:01}.

\subsection{Diffusion Operator}
\label{sec:4.4}
As shown in Fig.~\ref{fig:08}, each application of the Schr\"odinger's oracles is followed by the standard Grover diffusion operator $U_D$, acting on the entry label register. $U_D$ performs the conventional inversion about the mean, a key mechanism in amplitude amplification.

As shown in \eqref{eq:39}, what distinguishes our setting, however, is that the overall system evolves into a coherent superposition of two computational branches:
\begin{itemize}
\item[i)] a non-inverting branch, corresponding to the component $\ket{x} \otimes \ket{v_d^{\perp}}$
, in which the label has not undergone a phase flip, and
\item[ii)] an inverting branch, corresponding to 
$\ket{x} \otimes \ket{v_d}$, where the label has experienced a phase inversion, namely has been marked.
\end{itemize}
Here $\ket{x}$ denotes the label of the entry whose e-neighborhood contains the target address $\ket{v_d}$.
In the non-inverting branch, since no phase-inversion occurs, the subsequent application of $U_D$ leaves the amplitude distribution unchanged. In contrast in the inverting branch, the diffusion operator $U_D$ acts as in standard Grover: it amplifies the probability amplitude of hitting entry $\ket{x}$, while it suppresses those of the non-marked entries.

\subsection{Measurement}
\label{sec:4.5}
As shown in Fig.~\ref{fig:08}, the search is concluded upon the measurement of the entry label register, after a prescribed\footnote{This prescribed number can be computed by averaging over the optimal number of iterations for the two different branches.} number of iterations. This measurement collapses the superposition over the entry labels, yielding a specific outcome $\ket{x^*}$.

The probability of successfully identifying a hitting entry associated with the target address $\ket{v_d}$ benefits from two contributing factors: one coming from the non-inverting branch and the other from the inverting branch. Specifically, in the non-inverting branch, the amplitudes remain uniform, i.e., $\frac{1}{\sqrt{n_{\text{T}}}}$ and the measurement behaves like a uniform random choice over all the entries. Differently, in the phase-inverting branch, the amplitude of the hitting entry label associated to $\ket{v_d}$ is amplified toward unity, following standard Grover dynamics.

Overall, the probability of successfully measuring the marked entry $\ket{x}$ is enhanced by the Grover-amplified branch weighted by $\alpha$, while the no-inverting branch contributes with a uniform background distribution. Since $\alpha$ is proportional to $f(n_e)$, i.e., to the number of superposed addressed announced by each e-hop, $f(n_e)$ becomes pivotal in steering the algorithm’s success probability toward one.

Furthermore, while we assumed a single hitting entry $\ket{x}$ for simplicity, the expected number of such entries is much larger, on the order of $\log n_e$. This follows from our parametrization of the routing tables in terms of e-neighborhood size $k$ and number of anchors or tracking nodes in the Full-Anchor scheme), which is $\sqrt{n_e}$. Under this design, each quantum address appears redundantly in approximately $\log n_e$ distinct entries of a routing table. As a result, the probability of successfully detecting a hitting entry is further boosted by this built-in redundancy.

Although a comprehensive analytical and simulation-based characterization is left for future work, preliminary investigations show that the number of iterations required for address splitting grows only sub-linearly with the network size $n_e$, while the corresponding splitting delay depends on the choice of $f(n_e)$. In particular, increasing $f(n_e)$ raises the parameter $\alpha$, which simultaneously reduces the required number of iterations and drives the peak success probability closer to unity. Thus, adopting a poly-logarithmic scaling for $f(n_e)$, together with the inherent redundancy provided by approximately $\log{n_e}$ hitting entries, appears sufficient to ensure with high probability the discovery of a matching entry, even in quantum networks comprising millions of nodes.

\section{Discussion and Future Research Directions}
\label{sec:5}

This paper proposed a quantum-native architecture for the Quantum Internet, built on two synergistic pillars: the Entanglement-Defined Controller (EDC) and a novel quantum addressing scheme.

The EDC provides the foundational substrate by enabling a clear separation between the control and data planes, a necessary precondition for managing the inherently stateful nature of entanglement. The quantum addressing scheme completes this foundation, by embedding quantumness directly into node identifiers, thereby enabling the network to natively track and manipulate entanglement as a dynamic resource. This is key because entanglement is both ephemeral and non-local, requiring persistent in-network operations and continuous state awareness. Together, these pillars form the architectural foundation that enables scalability: compact and quantum-native routing protocols can be designed with provable performance guarantees, including constant entangling stretch and sublinear routing table size. In addition, the proposed address-splitting functionality -- powered by the Schrödinger’s oracle -- demonstrates the practical feasibility of manipulating superposed addresses, a crucial functionality for quantum-native control.\\
In a nutshell, our proposal is flexible, scalable and resilient. It supports both centralized and decentralized control infrastructure, adapting to the operational demands of large-scale quantum networks, as exemplified by the full-anchor
scheme.

In the following, we discuss several key aspects and implications of the proposed quantum-native functioning. For clarity and readability, each point is discussed individually, by highlighting specific aspects of the proposal and by providing context for rationale and future research directions.

\textbf{Compact Routing Design:} In 1977, Kleinrock and Kamoun published their pioneering paper \cite{KleKam-77} on hierarchical routing. Since then, hierarchical routing has been the foundation of both inter-domain and intra-domain routing techniques adopted in Internet, such as CIDR and OSPF/ISIS \cite{KriClaFal-07}. Kleinrock and Kamoun’s hierarchical approach was essentially the first \textit{name-dependent} routing scheme\footnote{In a nutshell, a \textit{name-dependent} routing scheme embeds some topological information within node addresses, which, thus, cannot be arbitrary. This topological information is, then, exploited to reduce the amount of information to be stored in each routing table.  Conversely, \textit{name-independent routing} works with topologically-agnostic node addresses. The differences between the two routing approaches are summarized in Table~\ref{tab:06}}. In the same paper, they were the first to analyze the stretch/routing-table size trade-oﬀ, by showing that the routing stretch produced by the hierarchical approach is satisfactory only for specific topologies. In other words, hierarchical routing is optimal by achieving shortest paths (namely, paths minimizing the route cost according to the adopted metric) with scalable routing table (routing tables that scale sublinearly with the number of nodes) for topologies with specific characteristics, such as trees or grids. But Internet does not satisfy this type of topologies. And indeed, analytical estimates show that applying hierarchical
routing to the Internet topology incurs a $\sim$ 15-times path length increase \cite{KriClaFal-07}.\\
\begin{table}[t]
    \renewcommand{\arraystretch}{1.4}
    \centering
    \begin{tabular}{p{0.29\columnwidth}|p{0.26\columnwidth}|p{0.26\columnwidth}}
        \hline
        \hline
        \textbf{property} & \textbf{name-dependent} & \textbf{name-independent} \\
        \hline
        address structure & topology-aware & topology-agnostic \\
        \hline
        routing table size & topology-dependent & sublinear (universal) \\
        \hline
        robustness to topology changes & limited & high \\
        \hline
        \hline
    \end{tabular}
    \caption{Name-dependent vs name-independent routing: concise comparative overview}
    \label{tab:06}
    \hrulefill
\end{table}
With this lesson learned from the classical Internet history, and by accounting for the unsuitability of topological-addresses for tracking entanglement as pioneered in \cite{CacIllCal-23}, we focus our attention on the design of \textit{universal quantum routing} schemes, namely, accordingly to the notation introduced in \cite{KriClaFal-07}, schemes that work correctly and satisfy promised scaling bounds on all graphs. The rationale for our design choice does not limit to the attractive feature of generality exhibited by universal routing protocols. But it accounts also for the conflicting preliminary results in terms of topology properties of quantum networks reported by recent literature \cite{BriCanCha-20,BriCanCav-21}, as well as for the absence of experimental studies on quantum network topologies (excepting extremely-small quantum networks such as \cite{ThoFeiChe-24}).\\
Last but not least, it is worthwhile to note that our proposal is the first compact routing protocol proposed in literature that achieves fully name-independent routing, i.e., that does not exploit an underlying name-dependent scheme \cite{KriClaFal-07}. This achievement is made possible by the design of the quantum addressing scheme, which encodes quantum properties directly into node identifiers. We believe that many other network functionalities could benefit similarly from quantum-native design principles.

\textbf{Worst-Case Baseline for Clustering:} The clustering of ESPs into the anchor set (or tracked set in the Full-Anchor scheme) has been performed using a flat, non-topology-aware partitioning (see Remarks~\ref{rem:07} and \ref{rem:08}). This provides a worst-case baseline, as no effort has been made to optimize the placement or connectivity of anchors/tracked ESPs. Our goal in this first analysis is to demonstrate that, even under such an uninformed clustering, the proposed schemes achieve coverage w.h.p. and constant-bounded entangling stretch. Thus, cluster optimizations are envisioned to definitively lead to even better performance. In this context, it is important to investigate whether the algorithms associated with the covering set should be executed distributively by the ESPs, either with or without EDC support, or remain entirely a responsibility of the EDC. Analyzing the trade-offs between these alternatives is crucial. In particular, it is necessary to understand how the computational efficiency evolves as the network scales, and to design appropriate countermeasures against potential performance bottlenecks, including the development of more efficient algorithms.

\textbf{Quantum Address Splitting \& Schr\"odinger’s Oracle:} The Address Splitting functionality proposed in this paper serves as a foundational proof of concept for extending classical forwarding operations to quantum-superposed identifiers. While potential optimizations, such as tuning the redundancy factor $\log_2{n_e}$ and the parameter $f(n_e)$, have been briefly discussed in Sec.~\ref{sec:4.5}, an important optimization dimension remains unexplored: the impact of anchor set (tracked set in the Full-Anchor scheme) structure on the effectiveness of address splitting. This opens the door to more efficient variants that could emerge from a joint optimization of the address splitting mechanism and clustering strategies. Exploring this synergy is a promising future direction.\\
The Schr\"odinger’s oracle $O_S$ designed to implement the quantum address splitting allows Grover’s algorithm to evolve in a coherent superposition of oracle-inverting and oracle-trivial dynamics. The presence or absence of the oracle’s action becomes itself a quantum degree of freedom, by introducing a new layer of coherence into the search process. We strongly believe this approach generalizes standard Grover search and opens new directions for exploring quantum algorithms operating in superpositions of operational regimes, enabled by quantum-superposed data structures. The potential advantages of this additional dimension of quantumness are reminiscent of the benefits observed in scenarios involving superpositions of quantum operations or tasks \cite{ChiDarPer-13,CalCac-20,SimCalIll-23,CalSimCac-23}. However, this remains an open question and warrants further investigation.

\textbf{Multipartite Entanglement:} The proposed quantum-native routing protocol considered only bipartite entanglement, as key communication resource. Yet, the full potential of the proposed quantum addressing scheme becomes even more evident in multipartite entanglement scenarios, where ESPs must address entire subsets of nodes. In such cases, quantum addressing and the tracking mechanisms proposed here are particularly well-suited, as they can significantly reduce routing table sizes that would otherwise scale super-linearly, without superposed addresses. Exploring this extension is part of our future work.

\textbf{Architecture Scalability \& Generality:} The proposed architecture integrates hierarchical principles with SDN principles, resulting in a two-tier structure that clearly separates the control and data planes. This separation simplifies network management by localizing decisions and reducing control overhead. The two-tier hierarchy enables scalable orchestration of entanglement resources and routing, while preserving flexibility and extensibility. Although our work focuses primarily on routing, the architectural foundation presented here is general and can support other quantum-native functions such as scheduling, fault tolerance, and resource provisioning. Thus, it provides a robust and adaptable foundation for the design and deployment of the Quantum Internet.\\
Within this architecture, a central role is played by the EDC and its operational model for information acquisition. A promising research direction is to investigate efficient techniques that allow the EDC to promptly gather the information needed for orchestrating entanglement resources and controlling the network, while limiting the associated overheads. We envision that progress in this direction will involve the hybridization of classical techniques from SDN controller design with the specific requirements imposed by quantum mechanics. Consequently, as also highlighted in the subsequent discussion, the interplay between classical and quantum infrastructures will be essential, by enabling eventually meaningful interactions between classical and quantum control planes. \\ 
Furthermore, the proposed architecture is not constrained by the assumption of a single, centralized EDC with complete network knowledge. Instead, multiple EDCs can coexist, either hierarchically organized or operating in a distributed fashion, with each EDC orchestrating a portion of the network. These EDCs can coordinate to share partial topological knowledge and enforce consistent entanglement resource policies, while still enabling local autonomy and scalability. In this perspective, the approach mirrors the design philosophy of distributed SDN controllers in classical networks, where control logic is logically centralized but physically distributed. An important line of future research is to understand how multiple EDCs can synchronize and coordinate efficiently as the network scales. This challenge is again closely tied to the broader question of quantum–classical coexistence and interplay. 

\textbf{Interplay with Classical Network Infrastructure:}
As emerged also in the architectural discussion above, no practical quantum network can operate in isolation from its classical counterpart. The coexistence of these two intertwined infrastructures greatly influences the overall scalability, as also discussed in \cite{IllCalMan-22,CacIll-22}. In fact, this interplay introduces additional scalability challenges, including signaling overhead and synchronization latency. In this work, our analysis has focused on establishing \textit{quantum-native} scalability, defined by performance guarantees -- such as constant entangling stretch and sublinear routing table size -- that address the fundamental challenges posed by the stateful and non-local nature of entanglement. 
A complete assessment of the hybrid system scalability not only constitutes a research direction in its own right, but it also requires the integration of these quantum-native metrics with their classical counterparts. Foundational works such as Santivanez \textit{et al.}~\cite{SanMcDSta-02} provide a rigorous framework for classical scalability. A promising line of future research is, therefore, to develop a unified framework that integrates classical and quantum scalability analyses, by combining classical overheads with the quantum-specific costs analyzed here. We believe that our results on compact quantum routing and constant entangling stretch provide a solid foundation upon which such hybrid analyses of quantum–classical scalability can be developed.  

\textbf{Simulator and Prototype Outlook:} 
This work has focused on theoretical derivations, since establishing a solid mathematical foundation has always been considered in literature \cite{KleKam-77} as both essential and preliminary for demonstrating the validity and scalability of a new approach. Numerical and, eventually, experimental validations represent the natural continuation of this research line. Unfortunately, at present, available quantum network simulators do not yet support quantum-native control functionalities, the very core of this work. Extending these tools to incorporate quantum-native control primitives -- including EDC orchestration, quantum addressing and manipulation of superposed addresses -- constitutes a very promising yet substantial research effort in its own right, and it is firmly part of our agenda.\\
Indeed, we have already initiated the design and development of a new quantum network simulator capable of natively supporting an inherently quantum control plane. In parallel, we are contributing to the international community efforts toward ``standardizing'' the concept of a quantum control plane and scaling quantumness in network architectures.\\
Beyond simulators, small-scale experimental prototypes will also be of great interest for assessing the robustness of the solutions under realistic network dynamics. Although such prototypes are likely to be limited in scale at first, they can provide crucial insights into error models, synchronization challenges, and the interplay with classical signaling. These practical lessons will, in turn, guide refinements of the theoretical model and the design of more efficient algorithms.\\
Taken together, these efforts are essential to complement the theoretical foundations laid in this paper, bridging the gap from formal proofs to simulation and, ultimately, experimental validation.

\section{Acknowledgment}
We are deeply indebted to John Day for generously sharing his authoritative insights on the historical origins of the Internet. As a pioneering contributor to the field since 1970 -- when his group at the University of Illinois was the 12th IMP installed but the 6th to go live on the ARPANET -- he offers a unique perspective as both a witness and a key architect. His later work in managing the development of the OSI reference model and contributing to naming, addressing, and upper-layer architecture further underscores the depth of his expertise. We feel truly honored to be in correspondence with him.

\begin{appendices}
\section{Proof of Lemma~\ref{lem:01}}
\label{app:02}
By denoting with $v_i$ and $v_d$ the ESPs exhibiting the worst-case entangling stretch in \eqref{eq:19}, we have two cases\footref{foot:03} with a entangling stretch larger than one: either \textit{Case II} or \textit{Case III}.

Let us focus first on \textit{Case III}. Accordingly, it results:
\begin{align}
    \label{app:02.1}
    w_\mathcal{R}(i,d) &= w\big( (i,l) \oplus (l,k) \oplus (k,d) \big)
\end{align}
By applying the axiomatic monotonic property of a metric given in \eqref{eq:14} to quantum path $(l,k)$, we can upper bound \eqref{app:02.1} as:
\begin{align}
    \label{app:02.2}
    w_\mathcal{R}(i,d) &= w\big( (i,l) \oplus (l,k) \oplus (k,d) \big) \leq \nonumber \\
    & \leq w\big( (i,l) \oplus (l,i) \oplus (i,k) \oplus (k,d) \big)
\end{align}
Clearly, it results $v_d \not \in N(v_i)$ -- otherwise \textit{Case I} would hold -- whereas $v_l \in N(v_i)$ by design. Thus, being $w(i,l) \leq w(i,d)$ from Def.~\ref{def:06} and by accounting for the axiomatic symmetry of a metric given in \eqref{eq:13}, we have:
\begin{equation}
    \label{app:02.3}
    w\big( (i,l) \oplus (l,i) \big) \leq w\big( (i,d) \oplus (d,i) \big)
\end{equation}
By extending both the entangling paths given in \eqref{app:02.3} with the common path $(i,k) \oplus (k,d)$ and by accounting for the right isotonicity property given in \eqref{eq:16}, we can upper bound \eqref{app:02.2} as:
\begin{align}
    \label{app:02.4}
    w_\mathcal{R}(i,d) &\leq w\big( (i,l) \oplus (l,i) \oplus (i,k) \oplus (k,d) \big) \nonumber \\
    &\leq w\big( (i,d) \oplus (d,i) \oplus (i,k) \oplus (k,d) \big)
\end{align}
By using the axiomatic monotonic property of a metric given in \eqref{eq:14} to the entangling path $(i,k)$, we can upper bound \eqref{app:02.4} as follows:
\begin{align}
    \label{app:02.5}
    w_\mathcal{R}(i,d) &\leq w\big( (i,d) \oplus (d,i) \oplus (i,k) \oplus (k,d) \big) \nonumber \\
    &\leq w\big( (i,d) \oplus (d,i) \oplus (i,d) \oplus (d,k) \oplus (k,d) \big)
\end{align}
We have that $v_i \not \in N(v_d)$ -- otherwise there would have been a corresponding entry for $v_d$ in $v_i$ quantum routing table, as pointed out with the remark after Lemma~\ref{lem:01}, and \textit{Case I} would hold -- while $v_k \in N(v_d)$ by design. Thus, being $w(d,k) \leq w(d,i)$ from Def.~\ref{def:06} and by accounting for the axiomatic symmetry of a metric given in \eqref{eq:13}, we have:
\begin{equation}
    \label{app:02.6}
    w\big( (d,k) \oplus (k,d) \big) \leq w\big( (d,i) \oplus (i,d) \big)
\end{equation}
By extending both the entangling paths given in \eqref{app:02.6} with the common path $(i,d) \oplus (d,i) \oplus (i,d)$ and by accounting for the left isotonicity property given in \eqref{eq:15}, we have the thesis, i.e.:
\begin{align}
    \label{app:02.7}
    w_\mathcal{R}(i,d) & \leq w\big( (i,d) \oplus (d,i) \oplus (i,d) \oplus (d,i) \oplus (i,d) \big) \nonumber \\
        & \eqdef w\left( \bigoplus_{5} (i,d) \right) \eqdef w_{\tilde{R}}(i,d).
\end{align}

As for \textit{Case II}, it follows straightforward from \textit{Case III} by denoting $v_k \eqdef v_k$ and by setting $w(i,l) = 0$. Indeed, in this particular case the entangling stretch exhibits an even tighter upper bound, i.e.:
\begin{align}
    \label{app:02.8}
    w_\mathcal{R}&(i,d)= w\big( (i,j) \oplus (j,d) \big) \leq w\big( (i,j) \oplus (j,i) \oplus (i,d) \big) \nonumber \\
    & \leq w\big( (i,d) \oplus (d,i) \oplus (i,d) \big) \eqdef w\left( \bigoplus_{3} (i,d) \right).
\end{align}

\end{appendices}

\bibliographystyle{ieeetr}
\bibliography{biblio.bib,biblio2.bib}

\begin{IEEEbiography}
[{\includegraphics[width=1in,height=1.25in,clip,keepaspectratio]{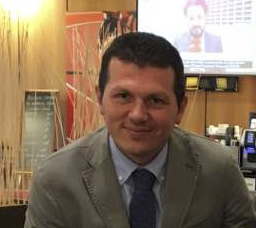}}]{Marcello Caleffi} \, (Senior Member, IEEE) is currently Professor of \textit{Advanced Quantum Networks} with the Department of Electrical Engineering and Information Technologies (DIETI), University of Naples Federico II, Naples, Italy, where he co-founded the Quantum Internet Research Group. His research has appeared in several premier IEEE Transactions and journals. He is the recipient of multiple awards, including the 2024 IEEE Communications Society Award for Advances in Communication and the 2022 IEEE Communications Society Best Tutorial Paper Award. He currently serves as Editor or Associate Editor for \textit{IEEE Transactions on Wireless Communications}, \textit{IEEE Transactions on Communications}, \textit{IEEE Transactions on Quantum Engineering}, \textit{IEEE Open Journal of the Communications Society}, and \textit{IEEE Internet Computing}. He has served as Chair and TPC Chair for several premier IEEE conferences. In 2017, he was appointed Distinguished Visitor Speaker by the IEEE Computer Society and was elected Treasurer of the IEEE ComSoc/VT Italy Chapter. In 2019, he was appointed as a member of the IEEE New Initiatives Committee by the IEEE Board of Directors, and in 2023, he was appointed as an IEEE ComSoc Distinguished Lecturer.
\end{IEEEbiography}

\begin{IEEEbiography}
[{\includegraphics[width=1in,height=1.25in,clip,keepaspectratio]{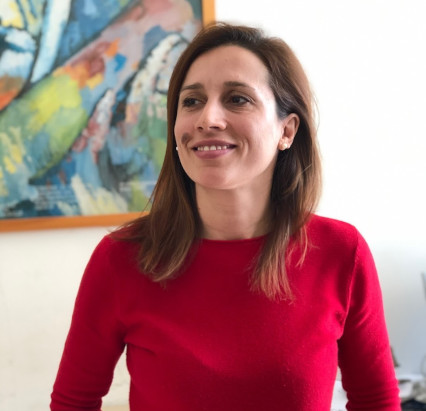}}]{Angela Sara Cacciapuoti} \, (Senior Member,
IEEE) is a Professor at the University of Naples Federico II, Italy, and co-founder of the Quantum Internet Research Group (www.quantuminternet.it). She is the PI of the ERC-Grant ``QNattyNet'', which aims to lay the foundations of a truly quantum-native Internet (www.qnattynet.quantuminternet.it). Prof. Cacciapuoti is a leading contributor to the theoretical and architectural foundations of quantum networking. Her work bridges quantum information theory, network design, and communication engineering, with a focus on enabling scalable quantum networks and hybrid quantum-classical architectures.
In recognition of her pioneering research, she has received several major international honors, including the ``2024 IEEE Communications Society Award for Advances in Communication'', the ``2022 IEEE ComSoc Best Tutorial Paper Award'', the ``2022 WICE Outstanding Achievement Award'', the ``2021 N2Women: Stars in Networking and Communications'', and the ``2023 IEEE ComSoc Distinguished Service Award (EMEA Region)''. She was also recently recognized as a \textit{Featured Author} on IEEE Xplore.
Prof. Cacciapuoti has been an IEEE ComSoc Distinguished Lecturer, delivering invited talks worldwide on Quantum Internet design and quantum network architectures. She currently serves as an Area Editor for IEEE Transactions on Communications and as a Senior Editor for the IEEE Journal on Selected Areas in Communications – Quantum Series. She is also on the editorial boards of npj Quantum Information, IEEE Transactions on Quantum Engineering, and IEEE Communications Surveys \& Tutorials. Previously, she served as Area Editor for IEEE Communications Letters (2019–2023), receiving the 2017 Exemplary Editor Award, and has held several IEEE leadership roles, including Vice-Chair and Publicity Chair of WICE and Treasurer of the IEEE Women in Engineering Affinity Group (Italy Section).
\end{IEEEbiography}

\end{document}